# Conditions for super-Eddington accretion onto the first black holes

Simone T. Gordon ,[★] Britton D. Smith , Sadegh Khochfar and Ricarda S. Beckmann
*Institute for Astronomy, University of Edinburgh, Royal Observatory, Edinburgh EH9 3HJ, UK*

**ABSTRACT**
Observations of supermassive black holes (BHs) at high redshift challenge our understanding of the evolution of the first generation of BHs in proto-galactic environments. One possibility is that they grow much more rapidly than current estimates of feedback and accretion efficiency permit. Following our previous analysis of super-Eddington accretion on to stellar-mass BHs in mini-haloes under no-feedback conditions, we now investigate whether this can be sustained when thermal feedback is included. We use four sets of cosmological simulations at sub-pc resolution with initial BH masses varying from $1 \times 10^3$ to $6 \times 10^4$ M$_\odot$, exploring a range of feedback efficiencies. We also vary the feedback injection radius to probe the threshold of numerical overcooling. We find that super-Eddington growth sustained of the order of ∼100 kyr is possible with weak thermal feedback efficiency in all environments and moderate efficiency for two of the BHs. Trans-Eddington growth is possible for a $3 \times 10^3$–$6 \times 10^3$ M$_\odot$ BH at moderate feedback efficiencies. We discuss the effectiveness of thermal feedback in heating the gas, suppressing accretion, and driving outflows at these parameter configurations. Our results suggest that super-Eddington growth may be possible in the presence of thermal feedback for BHs formed from the first stars.

## 1 INTRODUCTION

There are many unknowns concerning the first generation of black holes (BHs): their initial mass distribution, how they influence host haloes and galaxies, their growth pathways, and their feedback mechanisms. BHs observed in young galaxies by *JWST* are far more massive and abundant than theoretical models predicted (Madau & Rees 2001; Reed et al. 2019; Yang et al. 2020; Wang et al. 2021; Harikane et al. 2023; Greene et al. 2024; Maiolino et al. 2024c), indicating three possible scenarios: earlier formation, larger initial seed masses, or extended periods of super-Eddington accretion. A combination of these factors is likely needed to account for all the observed data.

Population III (Pop III) stars are thought to have formed between 100 and 250 Myr after the big bang at the beginning of the epoch of reionization (Klessen & Glover 2023). Star formation before this time is unlikely because metal-free gas requires longer cooling times than metal-enriched gas, and the high densities required for star formation in mini-haloes are orders of magnitude higher than the average density of the Universe. While a top-heavy initial mass function is expected due to the longer cooling times (Hirano et al. 2014), consistent near-Eddington growth is required for ∼$10^2$ M$_\odot$ BH remnants of Pop III stars to reach $10^9$ M$_\odot$ by $t \simeq 800$ Myr. Assuming that accretion is Eddington-limited, such a high duty cycle is unrealistic, given the suppressing effects of feedback from the accretion disc itself and ionizing radiation from supernovae (Trebitsch, Volonteri & Dubois 2019).

Alternative formation channels that give rise to larger seeds have been proposed, including direct-collapse BHs from supermassive stars (SMSs; Johnson et al. 2010; Agarwal et al. 2016; Regan & Downes 2018; Latif, Whalen & Khochfar 2022), stellar cluster mergers (Yajima & Khochfar 2016; Boekholt et al. 2018; Reinoso et al. 2023), and primordial BHs (Kawasaki, Kusenko & Yanagida 2012; Clesse & García-Bellido 2015; Carr & Silk 2018). The first two options solve the issue of small seeds, while primordial BHs could form as early as 1 s after the big bang with a >$10^2$ M$_\odot$ initial mass (Colazo, Stasyszyn & Padilla 2024). However, it is uncertain if the conditions required to form some of these massive seeds are prevalent enough to produce the observed number density of early Universe galaxy candidates hosting supermassive BHs (SMBHs), which may be higher than previously thought (Adams et al. 2022; Atek et al. 2023; Castellano et al. 2023; Harikane et al. 2023; Greene et al. 2024; Maiolino et al. 2024a). Direct-collapse BHs are expected to form from the near-isothermal collapse of a gravitationally unstable massive gas cloud at $T \simeq 10^4$ K, which results in a very high accretion rate and avoids major episodes of gas fragmentation (Inayoshi, Visbal & Haiman 2020) [though this has been called into question recently (Prole et al. 2024a)]. Mechanisms that suppress the formation of the coolant H$_2$ molecule must occur to keep the cloud warm and prevent fragmentation, such as dissociation from Lyman–Werner photons (Latif & Khochfar 2019), H$^-$ detachment from near-infrared and optical photons (Wolcott-Green & Haiman 2012), or collisional dissociation in shocked gas (Inayoshi & Omukai 2012). In some rare cases, rapid halo assembly can achieve a massive

---

[★] E-mail: simone.gordon@ed.ac.uk

atomically cooled halo via dynamical heating prior to star formation without suppressing H$_2$ abundances (Wise et al. 2019; Latif et al. 2022; Prole et al. 2024b). Stellar cluster formation requires very high star formation efficiencies and central cluster densities for metal-poor clouds to undergo runaway collapse into an intermediate-mass BH (IMBH) of $\simeq 10^4$ M$_\odot$ (Katz, Sijacki & Haehnelt 2015). Less dense stellar clusters could also form IMBHs via gas dynamical friction and hyper-Eddington growth (Ryu et al. 2016), though such efficient accretion is difficult to sustain.

The Eddington rate of spherically symmetric accretion on to a BH is defined as

$$\dot{M}_{\rm Edd} = \frac{4\pi G M_{\rm BH} m_{\rm p}}{\eta c \sigma_{\rm T}}, \quad (1)$$

where $M_{\rm BH}$ is the mass of the BH, $\eta$ is the radiative efficiency of the accreting matter, $m_{\rm p}$ is the mass of a proton, $\sigma_{\rm T}$ is the Thomson scattering cross-section, $G$ is the gravitational constant, and $c$ is the speed of light. This defines the maximum accretion rate a BH can achieve before radiation pressure prevents further accretion. A radiative efficiency of $\eta = 0.1$ is often adopted as it is consistent with the thin-disc accretion model for Schwarzschild BHs and has been validated on low-redshift quasars (Yu & Tremaine 2002). The thin-disc model describes a geometrically thin (thickness much smaller than radial extent) and optically thick (gas dense – radiation is emitted and absorbed multiple times before escaping) disc in a steady state, where the accretion rate is constant at all radii (Shakura & Sunyaev 1973; Soltan 1982). However, this model breaks down under conditions of very high accretion rates, which may be the case for high-redshift BHs at the start of their evolution. Theoretical works have proposed that BHs can grow at super-Eddington rates due to photon trapping in dense accreting material (Begelman 1979). When the inflow speed of gas exceeds the photon diffusion speed, radiation is trapped and advected towards the BH, limiting the observable luminosity to the Eddington luminosity regardless of the mass inflow rate. Paczynski & Abramowicz (1982) showed that a geometrically thick disc with equatorial accretion (as opposed to spherically symmetric) can support super-Eddington accretion rates without becoming dynamically unstable; the system can exceed the Eddington luminosity. However, later analytical studies and simulations found that at high accretion rates ($\dot{M}_{\rm BH} \sim 10^{-3} \dot{M}_{\rm Edd}$ M$_\odot$ yr$^{-1}$), radiatively inefficient accretion streams become unstable, leading to significant outflows in the polar directions and limiting the mass reaching the BH (Quataert & Gruzinov 2000; Igumenshchev, Narayan & Abramowicz 2003; Blandford & Begelman 2004). The photon-trapping effect has been integrated into accretion disc models and direct radiation hydrodynamics simulations (RHD), some of which include general relativistic effects and magnetic fields (GRRMHD). It has been found that radiative efficiency decreases modestly with increasing accretion, reaching $\eta = 0.01$ at super-Eddington rates (Takahashi & Ohsuga 2015; Ryan et al. 2017) [though higher efficiencies for similar accretion rates are still found (Fragile et al. 2018)] and that magnetic fields can both drive and hinder accretion under different conditions (Mishra et al. 2022). The GRRMHD simulations presented in Sadowski et al. (2015) found $\eta < 0.01$ and $\dot{M}_{\rm BH} \sim 10^3 \dot{M}_{\rm Edd}$ M$_\odot$ yr$^{-1}$ in the case of a magnetically arrested disc, where strong magnetic fields dominate the inflow of gas.

There have been multiple candidates for super-Eddington SMBH accretion reported in recent times (Du et al. 2018; Yue et al. 2023; Bhatt et al. 2024; Jin et al. 2024; Maiolino et al. ; Suh et al. 2024; Yang et al. 2024), but there may be even more if certain assumptions are re-evaluated. For example, King (2024) argues that high-redshift quasars might have much lower BH masses than currently estimated if their luminosities are due to beamed, super-Eddington accretion. Likewise, BHs can achieve the same luminosity as SMBHs at lower masses if the accretion process keeps their spin and radiative efficiency low (King, Pringle & Hofmann 2008), removing the need for massive BH (MBH) seeds. This can occur if the accretion on to the BH is chaotic, involving successive accretion events with uncorrelated angular momenta. This perspective aligns with the scale-free nature of BH accretion processes observed in ultraluminous X-ray sources and suggests that further observational tests are necessary to confirm these implications (King & Lasota 2024). In a similar vein, Lupi et al. (2024a) question mass measurements of high-redshift MBHs observed by *JWST*, finding that uncertainties may be higher than previously thought and propose that some MBHs are smaller and accreting more efficiently than current estimates. This would relieve certain observational tensions; for instance, aligning these high-redshift observations closer to the local $M_*$–$M_{\rm BH}$ relation. Non-Eddington-limited accretion models could also solve the IMBH gap. Piana, Pu & Wu (2024) argue that BHs in the IMBH mass range ($10^4$–$10^6$ M$_\odot$) experience rapid growth due to hyper-Eddington accretion episodes, resulting in masses increasing by one to two orders of magnitude within just 20 Myr. This implies that IMBHs are not observed because they do not exist for long; super-Eddington growth causes them to quickly transition to SMBHs.

In our prior study, Gordon et al. (2024) (G24 henceforth), we showed that a direct-collapse BH born at the centre of a dark matter mini-halo at $z \sim 20$ could sustain accretion at super-Eddington rates in the absence of feedback. We now investigate the impact of varying degrees of thermal feedback on this kind of BH in a similar set-up. The paper is organized as follows. Section 2 details the set-up of our simulations. The results are presented in Section 3. Section 4 contains a comprehensive discussion of the results and comparisons with other works. Section 5 highlights some caveats and limitations of our study. Finally, we summarize our conclusions in Section 6.

## 2 THE SIMULATIONS

We investigated the impact of thermal feedback on super-Eddington accretion in early Universe BHs using sub-pc resolution cosmological simulations. We allowed a single stellar-mass BH born in a dark matter mini-halo to evolve unencumbered by feedback for a certain period of time, and then resimulated with thermal feedback activated before a significant accretion event. We did this for four accretion events, resulting in four simulations with initial BH masses ranging from $10^3$ to $10^5$ M$_\odot$. We use the simulation code ENZO (Bryan & Enzo Collaboration 2014; Brummel-Smith et al. 2019) to generate all data analysed in this work. ENZO is a hydrodynamical, block-structured adaptive mesh refinement (AMR) + *N*-body cosmological simulation code that is capable of achieving arbitrarily high spatial and temporal resolution. Our simulations are designed to follow the collapse of a metal-free gas cloud within a single cosmological halo, the same set-up as in G24. We initialize the simulations with a 500 kpc h$^{-1}$ comoving box at $z = 180$ using the MUSIC (Hahn & Abel 2011) initial conditions generator with the *Wilkinson Microwave Anisotropy Probe 7* best-fitting cosmological parameters, $\Omega_{\rm m} = 0.266$, $\Omega_\lambda = 0.732$, $\Omega_{\rm b} = 0.0449$, $H_0 = 71.0$ km s$^{-1}$ Mpc$^{-1}$, $\sigma_8 = 0.801$, and $n_{\rm s} = 0.963$ (Komatsu et al. 2011), the Hu & Eisenstein (1999) transfer function, and second-order Lagrangian perturbation theory. We run all simulations with a root grid of $128^3$ dark matter particles and cells.

In order to identify two suitable haloes for BH formation, we run dark matter-only simulations to $z = 10$ and locate a halo with total mass $\simeq 10^7$ M$_\odot$ in two sets of initial conditions using the ROCKSTAR halo finder (Behroozi et al. 2013). We then resimulate from $z = 180$ with additional four levels of telescoping refinement around the target halo, adding baryons in the last iteration. The high-resolution region is a rectangular prism that contains all dark matter particles destined to reside within three virial radii of the target halo at $z = 10$. These particles are designated as 'must-refine particles', ensuring that AMR is exclusively applied within this zone (Simpson et al. 2013). Prior to AMR, the high-resolution region has an effective resolution of $2048^3$, corresponding to a comoving spatial resolution of 0.244 kpc h$^{-1}$, a baryon mass resolution of 0.259 M$_\odot$, and a dark matter mass resolution of 1.274 M$_\odot$. It has dimensions of $48 \times 56 \times 60$ cells, or a comoving volume of $11.7 \times 13.7 \times 14.6$ kpc$^3$ h$^{-3}$.

**Table 1.** Summary of each accretion event. Events 1 and 2 occur in Halo 1 and events 3 and 4 occur in Halo 2. The columns from left to right represent as follows: $z_{init}$ and $t_{init}$(Myr) refer to the redshift and age at which the BHs are resimulated with thermal feedback activated, respectively; $M_{BH}$ (M$_\odot$) represents the BH mass at the start of the resimulated period; $M_{disc}$ (M$_\odot$) is the initial mass of the disc surrounding the BH; $dx_{min}$ (pc) and $dx_{max}$ (pc) indicate the minimum and maximum spatial resolution (cell width) of the simulation, respectively; $t_{end}$ (Myr) refers to the age of the BH at the end of the accretion event; $\Delta M_{BH}$ (M$_\odot$) indicates the total mass gained by the BH in the no-feedback simulation during the event; and $\langle \Delta \dot{M}_t / \dot{M}_{t-x/t+x} \rangle$ represents the average increase in accretion rate during the event relative to the accretion rate $t = 200$ kyr before and after the event (magnitude of the event).

| Event | $z_{init}$ | $t_{init}$(Myr) | $M_{BH}$ (M$_\odot$) | $M_{disc}$ (M$_\odot$) | $dx_{min}$ (pc) | $dx_{max}$ (pc) | $t_{end}$ (Myr) | $\Delta M_{BH}$ (M$_\odot$) | $\langle \Delta \dot{M}_t / \dot{M}_{t-x/t+x} \rangle$ |
|---|---|---|---|---|---|---|---|---|---|
| 1 | 25.95 | 2.50/2.70 | $3.2 \times 10^3$ | $5 \times 10^3$ | $3.12 \times 10^{-3}$ | $1.43 \times 10^{-2}$ | 2.80 | 627 | x2 |
| 2 | 22.48 | 31.70 | $6.0 \times 10^4$ | $6 \times 10^4$ | $1.43 \times 10^{-2}$ | $1.43 \times 10^{-2}$ | 32.00 | 2352 | x4 |
| 3 | 19.20 | 0.50 | $1.1 \times 10^3$ | – | $2.07 \times 10^{-3}$ | $8.30 \times 10^{-3}$ | 0.80 | 1266 | x3 |
| 4 | 18.79 | 6.50/6.90 | $6.6 \times 10^3$ | $5 \times 10^3$ | $4.24 \times 10^{-3}$ | $4.24 \times 10^{-3}$ | 7.00 | 1166 | x7 |

AMR occurs only in the high-resolution zoom-in region that is traced by the must-refine particles. These tracer particles indicate that their host cell is destined to reside in the target halo at the end of the simulation. Cells are split by factors of two in each dimension when one of the following conditions is met:

i. The dark matter mass within a grid cell is greater than four times the initial mass (i.e. when more than four dark matter particles are in the same cell).
ii. The gas mass within a grid cell is greater than four times the mean baryon mass per cell on the root grid multiplied by a factor, $2^{-0.2L}$, where $L$ is the refinement level.
iii. The local Jeans length is resolved by less than 32 or 64 cells.

Properties of each of the two haloes pre-BH formation are shown in table 3 and fig. 1 of G24, which we summarize here. The BH is seeded in Halo 1 at $z \simeq 26$ with virial mass of $3 \times 10^5$ M$_\odot$, and at $z \simeq 19$ with virial mass of $1 \times 10^6$ M$_\odot$ in Halo 2. Details on the seeding mechanism can be found in Section 2.2. Both are close to the minimum baryonic mass a virialized cloud must have to cool sufficiently for the formation of Pop III stars (Tegmark et al. 1997; Abel, Bryan & Norman 2002). This distinguishes these structures as mini-haloes as opposed to young galaxies (Greif et al. 2008).

Within the dark matter mini-halo, we simulate the evolution of a single stellar-mass BH, allowing it to grow without feedback for a period. We then resimulate at $t_{init}$ with thermal feedback enabled just prior to a major accretion event, defined simply as short periods ($<1$ Myr) of elevated accretion (a factor of a few greater than in the $\sim 200$ kyr before and after) identified in the simulations without feedback. This process is repeated for four accretion events across two haloes, producing four distinct simulations with initial BH masses between $10^3$ and $10^5$ M$_\odot$. We summarize these simulations in Table 1.

### 2.1 Cooling and heating

We use the GRACKLE non-equilibrium chemistry network (Smith et al. 2017), which follows the interactions of the nine dominant chemical species in primordial gas: H, H$^+$, H$^-$, $e^-$, He, He$^+$, He$^{++}$, H$_2$, and H$_2^+$. The radiative losses from atomic and molecular line cooling, Compton cooling, and heating of free electrons by cosmic microwave background (CMB) photons are appropriately treated in the optically thin limit. This network includes H$_2$ formation from the H$^-$ and H$_2^+$ channels, three-body formation according to the Glover (2008) prescription, H$_2$ rotational transitions, chemical heating, and collision-induced emission [important in $n \geq 10^{14}$ cm$^{-3}$ gas (Ripamonti & Abel 2004)]. The simulations presented here reach densities of $n > 10^9$ cm$^{-3}$ in which regime the three-body H$_2$ formation channel has been shown to dominate the cooling (Abel et al. 2002; Turk et al. 2010); hence, its inclusion is essential to accurately capture the gas evolution on these scales.

### 2.2 Star/black hole particle formation

We use the ENZO module SmartStars to simulate particle formation and accretion within the high-resolution region. This module was first created by Regan & Downes (2018) to simulate the evolution of supermassive Pop III stars in atomic cooling haloes. We have adapted it for our purposes, the details of which can be found in G24. We summarize the key points here. A Pop III star particle is inserted in a grid cell where the following conditions are met:

i. The proper baryon number density exceeds $10^6$ cm$^{-3}$.
ii. The gas flow is convergent.
iii. The molecular hydrogen mass fraction $f_{H2} \equiv (\rho_{H2} + \rho_{H2+})/\rho_b$ exceeds $5 \times 10^{-4}$, where $\rho_{H2}$, $\rho_{H2+}$, and $\rho_b$ are the neutral molecular hydrogen, singly ionized molecular hydrogen, and the total baryon densities, respectively.
iv. There is at least twice the mass of the particle in gas within a sphere of radius $r \simeq 1$ pc.

We implemented a formation model based on Wise et al. (2012) in which the star is initialized with its final mass as specified by the user, and stellar accretion is turned off. This approach necessitates the fourth formation criterion, which ensures the conservation of energy-momentum by instantaneously removing the star's entire mass from the grid. The star particles in this study immediately transition to BHs of equal mass as they all exceed 260 M$_\odot$ (the maximum stellar mass for a pair-instability supernova death, which completely destroys the progenitor) and have zero metallicity (Woosley, Heger & Weaver 2002).

## 2.3 Black hole accretion

We employ the Bondi–Hoyle–Lyttleton (BHL) sub-grid model:

$$\dot{M}_{\rm BHL} = \alpha \frac{4\pi G^2 M_{\rm BH}^2 \rho_\infty}{(c_{s,\infty}^2 + v_\infty^2)^{\frac{3}{2}}}, \quad (2)$$

where $\alpha$ is a dimensionless factor, $G$ is the gravitational constant, and $M_{\rm BH}$ is the mass of the BH (Hoyle & Lyttleton 1939; Bondi 1952). The input variables are the density of the gas $\rho_\infty$, the sound speed $c_{s,\infty}$, and the velocity $v_\infty$, where the latter is with respect to the BH velocity. They are all mass- and Gaussian kernel-weighted. The BH is allowed to accrete gas from within the accretion radius, defined as

$$r_{\rm acc} = \max(R_{\rm HL}, {\rm d}x), \quad (3)$$

where $R_{\rm HL}$ is Hoyle-Lyttleton (HL) scale radius of BHL accretion,

$$R_{\rm HL} = \frac{2GM_{\rm BH}}{v_\infty^2}, \quad (4)$$

and d$x$ is the width of the cells in the high-resolution region. Typically, $R_{\rm HL}$ does not exceed twice d$x$. There is a 75 per cent cap on the quantity of gas that can be removed from a cell at each time-step. Further details can be found in G24 (section 2.5.2).

## 2.4 Black hole thermal feedback

The precise mechanism by which energy emitted from a BH is coupled to the surrounding medium is as yet unknown (Booth & Schaye 2009). In our model, BHs inject a fixed fraction of the rest-mass energy of the gas they accrete into the surrounding medium. The feedback is implemented thermally; that is, energy is deposited into the surrounding gas by increasing its internal energy. The fraction of the accreted rest-mass energy that is injected is assumed to be independent of both the environment and the accretion rate. The amount of energy returned by a BH to its surrounding medium in a time-step d$t$ is given by

$$dE_{\rm BH\text{-}feed} = \epsilon_f \epsilon_r \dot{M}_{\rm BH} c^2 \, dt, \quad (5)$$

where $\dot{M}_{\rm BH}$ is the accretion rate on to the BH and $c$ is the speed of light. The radiative efficiency factor $\epsilon_r$ refers to the fraction of accreted matter that is released as radiation, and it is calculated based on $\eta$ (as in equation 1):

$$\epsilon_r = \frac{\eta}{1-\eta}. \quad (6)$$

For an optically thick and geometrically thin accretion disc, $\eta = 0.1$; hence, $\epsilon_r = 0.11$. However, as mentioned in Section 1, this assumption may not apply in the case of geometrically thick discs accreting at super-Eddington rates. We set $\epsilon_r \in [0.01, 0.1]$, which correspond to $\eta = 0.11$ and $\eta = 0.01$, respectively. The feedback efficiency parameter $\epsilon_f$ is the efficiency with which the radiated luminosity from the disc couples thermodynamically to the surrounding gas; we adopt values $\epsilon_f \in [0.05, 0.001, 0.0001]$. The range of overall accretion efficiency explored is $1 \times 10^{-5} \leq \epsilon_f \epsilon_r \leq 5 \times 10^{-3}$. Feedback efficiency can be reduced by numerical overcooling if the thermal energy is distributed over too much gas, causing smaller temperature increases and rapid radiative losses (Dalla Vecchia & Schaye 2012; Wang et al. 2018). This motivates us to also vary the radius of the feedback sphere, $r_{\rm fb} \in [5{\rm d}x, 7{\rm d}x, 10{\rm d}x]$.

In Di Matteo, Springel & Hernquist (2005), Springel, Di Matteo & Hernquist (2005), Sijacki et al. (2007), and Di Matteo et al. (2008), $\epsilon_f = 0.05$ to ensure that the merging SMBHs simulated grow according to the observed $M_{\rm BH}$–$\sigma$ relation, where $\sigma$ is the velocity dispersion of stars in the galactic bulge. A value of $\epsilon_f = 0.15$ is used in Booth & Schaye (2009) and Oppenheimer et al. (2019), where it is noted that 'this BH model is highly successful at reproducing the observed soft and hard X-ray luminosity functions of AGN'. In Taylor & Kobayashi (2014), $\epsilon_f = 0.25$ is shown to reproduce the observed $M_{\rm BH}$–$\sigma$ relation up to $z = 9$ for a BH of seed mass of $10^3 \, h^{-1} \, {\rm M}_\odot$. They find that lower values make BHs more massive at a given galaxy mass. They state that with a larger $\alpha$ (25–100) and smaller seed mass of $10^2 \, {\rm M}_\odot$, the cosmic star formation rate history and $M_{\rm BH}$–$\sigma$ relation are still fairly well reproduced, but the sizes of galaxies tend to be even larger than observed. In table 2 of Wang et al. (2018), a summary of the feedback efficiency parameters used in large modern simulation suites – TK, ILLUSTRIS, ILLUSTRISTNG, and EAGLE – is given. The product $\epsilon_f \epsilon_r$ differs by a factor of up to 2.5 in the simulations compared. However, unlike some of the above examples, this work focuses on BHs for which no observed scaling relations have been probed. Furthermore, our simulations can resolve down to the HL radius (equation 4) scale and amply resolve the Bondi radius (same as equation 4 but with $c_{s,\infty}$ in the denominator instead). The other works mentioned have not done this and so do not explicitly resolve the gas dynamics and dynamical friction relevant to accretion. See the appendices of G24 for further details on the resolution of the Bondi and HL radii throughout the evolution of our no-feedback simulations.

While the $M_{\rm BH}$–$\sigma$ relation has been shown to be redshift-independent (Ferrarese et al. 2001; Wyithe & Loeb 2003), the efficiency of radiation gas coupling depends on many factors: the BH mass, its mode of accretion, and the local gas properties. It is difficult to extrapolate the value for $\epsilon_f$ in the above simulations to our set-up. In less dense environments with lower accretion rates, the feedback efficiency might be lower because a smaller fraction of the accreted energy would be needed to influence the halo's dynamics. Conversely, if the accretion rate is near the Eddington limit and the radiative efficiency is high, a higher value of $\epsilon_f$ might be needed to accurately model the feedback effects.

The injected energy d$E_{\rm BH\text{-}feed}$ (see equation 5) is computed at each time-step and is deposited into a sphere with a minimum radius of five times the width of the BH host cell (5d$x$), which corresponds to about 524 cells. We also vary this radius to $5 \, {\rm d}x \leq r_{\rm fb} \leq 20 \, {\rm d}x$. These values were chosen such that the sphere fits within the fully resolved square feedback zone around the particle ($11^3 \times {\rm d}x$). It was necessary to implement an energy-budgeting mechanism for the thermal feedback. The environment from which these BHs are accreting is initially much denser than the BHs that formed from a stellar collapse or in the wake of a supernova. Since the energy released is directly proportional to the accretion rate, the BH thermal feedback calculated from equation (5) is extremely high and the Riemann solver slows significantly as the time-step shrinks. We therefore impose a limit on the specific energy (erg g$^{-1}$) of the gas per cell based on a maximum temperature $T_{\rm max}$:

$$e_{\rm max} = \frac{k_{\rm B} T_{\rm max}}{(\gamma - 1) \mu m_{\rm H}}, \quad (7)$$

where $k_{\rm B}$ is the Boltzmann constant, $m_{\rm H}$ is the mass of one hydrogen atom, $\gamma$ is the adiabatic index of the gas, and $\mu$ is the mean molecular weight of the gas. $T_{\rm max} = 1 \times 10^8$ K, $\mu = 0.58$ for fully ionized gas, and $\gamma$ is calculated on the fly. If the specific energy in a cell exceeds $e_{\rm max}$, the surplus energy is stored in a variable $E_{\rm store}$ and injected as feedback at a later time-step when the energy threshold is no longer exceeded. This energy budgeting scheme ensures that mass-energy is conserved while releasing the thermal feedback energy at a manageable rate for the simulation.

## 3 RESULTS

### 3.1 Simulation suite overview

We study the impact of a range of feedback efficiencies on four distinct accretion events, defined simply as short periods (<1 Myr) of elevated accretion (a factor of a few greater than in the ∼200 kyr before and after) identified in the simulations without feedback. The key properties of each accretion event are summarized in Table 1, while their associated simulations are detailed in Table 2. With each event, we focus on different facets of the problem. Event 1 is resimulated across two resolution levels and initiated from two different starting points $t_{init}$, henceforth referred to as 'early' and 'late' starts, allowing us to explore the impact of resolution and feedback duration. Event 2 has the most massive BH at $M_{BH} = 6 \times 10^4$ M$_\odot$, resembling an MBH more closely than a stellar seed mass BH, and also has the largest range of feedback parameter combinations (both efficiency and injection radius). Accretion Event 3 has the smallest BH, which has not yet developed a sub-pc scale disc and is resimulated over three refinement levels, enabling us to explore the impact of feedback on disc formation and its dependence on spatial resolution. Finally, Event 4 is the most intense and distinct, averaging a rate seven times higher within <0.1 Myr than the intervals before and after the event. The magnitude of this event makes it a good candidate to test subtle differences between feedback efficiencies.

The simulations are labelled according to the following convention: [event number: 1–4][early/late start: E/L][simulation resolution level]-$\epsilon_r$[radiative efficiency]-$\epsilon_f$[feedback efficiency]-[number of cells in feedback injection sphere radius]dx. For example, 1E14-$\epsilon_r$0.1-$\epsilon_f$0.05-5dx refers to an early start Event 1 simulation at resolution level 14 with $\epsilon_r = 0.1$, $\epsilon_f = 0.05$, and $r_{fb} = 5$d$x$.

### 3.2 Disc fragmentation promotes accretion

Disc fragmentation plays a key role in facilitating accretion on to the BH. We analyse the fragmentation of the gas to help characterize the impact of progressively more efficient thermal feedback using a clump-finding algorithm in a manner similar to Smith et al. (2009). Clumps are defined as masses enclosed by the lowest isodensity surface around a local density maximum. The algorithm first defines density contours within a 4 pc radius sphere centred on the BH particle. Starting with a single encompassing contour, subsequent iterations refine these contours by incrementally increasing the minimum density by a factor of 2, thereby segmenting the data into parent and child clumps. A clump is considered valid if it has negative total energy. That is, the gas kinetic energy, thermal energy, and radiative losses are summed together over a free-fall time and subtracted from the gravitational potential energy of the gas and collisionless particle system. Furthermore, we are primarily concerned with clumps that may be accreted by the BH. Clumps of gas with high angular momentum, such as a disc structure, are unlikely to do so. Hence, we clean the clump data to remove overmassive clumps very close to the BH, verifying that they correspond to disc structures by reviewing density projections.

Comparing the high-resolution Event 1 simulation 1E16-no-feedback to its lower resolution counterpart 1E14-no-feedback in Fig. 1, we observe the former accreting almost eight times as much as the latter during the whole period. This discrepancy is attributable to the disc structure; while the disc in 1E14-no-feedback remains smooth, the disc in 1E16-no-feedback fragments into clumps as early as $t = 2.30$ Myr. These clumps are then incorporated into the inner disc that feeds the BH, driving the accretion event, as seen in Fig. 2. Despite 1E14-no-feedback starting with a higher BH mass of $M_{BH} = 4.3 \times 10^3$ M$_\odot$ at 2.70 Myr compared to $M_{BH} = 3.6 \times 10^3$ M$_\odot$ in 1E16-no-feedback, the clumpy accretion in the latter allows it to surpass the mass of 1E14-no-feedback by 3.20 Myr. This illustrates the greater efficiency of episodic accretion in contrast to smooth accretion.

Furthermore, fragmentation helps mitigate the effects of thermal feedback. In the lower resolution simulation 1L14-$\epsilon_r$0.01-$\epsilon_f$0.001-5dx, feedback suppresses accretion more effectively, with accretion immediately plummeting to trans-Eddington rates. In contrast, the higher resolution 1L16-$\epsilon_r$0.01-$\epsilon_f$0.001-5dx allows $f_{Edd} > 10$ super-Eddington accretion to persist. This trend holds for higher efficiency accretion also; 1L14-$\epsilon_r$0.1-$\epsilon_f$0.05-5dx falls rapidly to $f_{Edd} < 1 \times 10^{-3}$, whereas 1L16-$\epsilon_r$0.1-$\epsilon_f$0.05-5dx declines more gradually. The difference in sensitivity to feedback between resolutions is due to two factors. The physically larger feedback injection radius in lower resolution simulations (the radius is a multiple of cell width d$x$) diffuses the energy over a greater volume. This can lead to an overestimation of feedback effects, as the injected thermal energy is spread more evenly and quickly throughout the simulation volume. Secondly, the denser clumps present in 1E16-no-feedback can radiate away energy, further reducing feedback's effectiveness in halting accretion.

It should be noted that fragmentation before disc formation can impede growth. In Halo 2, the gas cloud did not radially collapse to one point, but rather broke off into two clumps. As a result, there is an interaction between a small compact proto-disc around the BH and a self-gravitating cluster of clumps less than a parsec away. The cluster pulls gas from the proto-disc and disrupts its growth until the two structures eventually coalesce, leading to Accretion Event 4 between $t = 0.5$ and 0.8 Myr in the no-feedback run. When thermal feedback is activated at the beginning of this coalescence, the proto-disc is depleted entirely at mid to maximum feedback efficiencies. With no reservoir of dense gas to fuel it, the BH grows little thereon.

### 3.3 The impact of feedback on the disc

Feedback destroys clumps and homogenizes the gas in the vicinity of the BH. Fig. 3 compares the evolution of clump properties in 1E16-no-feedback and 1L16-$\epsilon_r$0.1-$\epsilon_f$0.05-5dx (most efficient feedback) during Accretion Event 1. The figure is divided into two panels for each case, with the top panels showing the clump distribution, disc radius, and BH mass, and the bottom panels illustrating the evolution of the total/max clump mass, disc mass, and clump count within 0.5 pc of the BH. In the no-feedback simulation (left), clumps are more densely distributed, and there is often a massive 101–200 M$_\odot$ clump (blue point) close to the BH (within 0.1 pc). In the max feedback simulation (right), the clumps are fewer and farther apart, especially towards the later stages where there are no clumps whatsoever within 0.1 pc of the BH. The disc radius (defined as the minimum volume that encloses gas of average density $n \geq 1 \times 10^6$ cm$^{-3}$) also reduces throughout the accretion period in the feedback simulation but remains relatively stable without feedback. The bottom panel shows a slight upward trend in total and max clump mass and clump number in the no-feedback case, in contrast with a steep decline in the presence of feedback. The plateau in the BH mass (grey line, top panel) in the latter case corresponds with the trough in three of the clump statistics, suggesting that without clumps the BH does not grow. Interestingly, while the total clump mass in the no-feedback run is about double that of the with-feedback run on average, there is little difference in the maximum clump mass. This suggests that the largest clumps can withstand the impact of feedback better than less massive clumps. Feedback heats the gas, which increases the Jeans mass and destabilizes smaller clumps, making them gravitationally unbound before larger ones. We can conclude that strong feedback expels gas and clumps from around the BH, breaks clumps apart, and thwarts the formation of new clumps in Accretion Event 1.

**Table 2.** Summary of the simulations across two sets of initial conditions (haloes) and four accretion events. From left to right, the columns represent (1) the simulation name; (2) the initial mass of the BH in solar masses; (3) the time at which the simulation was restarted with feedback; (4) the radiative and accretion feedback parameters; (5) the resolution of the accretion region in units of proper parsecs; (6) the corresponding resolution level of the ENZO simulation; and (7) the approximate end time of the accretion event.

| | Resimulated from accretion event with BH thermal feedback | | | | | | |
|---|---|---|---|---|---|---|---|
| Simulation | BH mass ($M_\odot$) | Start (Myr) | $\epsilon_r$ | $\epsilon_f$ | dx (pc) | Level | Event end (Myr) |
| 1E16-no-feedback | $3.23 \times 10^3$ | 2.50 | None | None | $3.12 \times 10^{-3}$ | 16 | 2.80 |
| 1E16-$\epsilon_r$0.1-$\epsilon_f$0.05-5dx | $3.23 \times 10^3$ | 2.50 | 0.1 | 0.05 | $3.12 \times 10^{-3}$ | 16 | 2.80 |
| 1E16-$\epsilon_r$0.1-$\epsilon_f$0.05-7dx | $3.23 \times 10^3$ | 2.50 | 0.1 | 0.05 | $3.12 \times 10^{-3}$ | 16 | 2.80 |
| 1E16-$\epsilon_r$0.1-$\epsilon_f$0.05-10dx | $3.23 \times 10^3$ | 2.50 | 0.1 | 0.05 | $3.12 \times 10^{-3}$ | 16 | 2.80 |
| 1E16-$\epsilon_r$0.01-$\epsilon_f$0.001-5dx | $3.23 \times 10^3$ | 2.50 | 0.01 | 0.001 | $3.12 \times 10^{-3}$ | 16 | 2.80 |
| 1L16-$\epsilon_r$0.1-$\epsilon_f$0.05-5dx | $3.60 \times 10^3$ | 2.70 | 0.1 | 0.05 | $3.12 \times 10^{-3}$ | 16 | 2.80 |
| 1L16-$\epsilon_r$0.01-$\epsilon_f$0.05-5dx | $3.60 \times 10^3$ | 2.70 | 0.1 | 0.05 | $3.12 \times 10^{-3}$ | 16 | 2.80 |
| 1L16-$\epsilon_r$0.01-$\epsilon_f$0.001-5dx | $3.60 \times 10^3$ | 2.70 | 0.01 | 0.001 | $3.12 \times 10^{-3}$ | 16 | 2.80 |
| 1L14-$\epsilon_r$0.1-$\epsilon_f$0.05-5dx | $3.60 \times 10^3$ | 2.70 | 0.1 | 0.05 | $1.43 \times 10^{-2}$ | 14 | 2.80 |
| 1L14-$\epsilon_r$0.01-$\epsilon_f$0.05-5dx | $3.60 \times 10^3$ | 2.70 | 0.01 | 0.05 | $1.43 \times 10^{-2}$ | 14 | 2.80 |
| 1L14-$\epsilon_r$0.01-$\epsilon_f$0.001-5dx | $3.60 \times 10^3$ | 2.70 | 0.01 | 0.001 | $1.43 \times 10^{-2}$ | 14 | 2.80 |
| 2E14-no-feedback | $6.025 \times 10^4$ | 31.70 | None | None | $1.43 \times 10^{-2}$ | 14 | 32.10 |
| 2E14-$\epsilon_r$0.01-$\epsilon_f$0.05-5dx | $6.025 \times 10^4$ | 31.70 | 0.1 | 0.05 | $1.43 \times 10^{-2}$ | 14 | 32.10 |
| 2E14-$\epsilon_r$0.01-$\epsilon_f$0.05-7dx | $6.025 \times 10^4$ | 31.70 | 0.01 | 0.05 | $1.43 \times 10^{-2}$ | 14 | 32.10 |
| 2E14-$\epsilon_r$0.01-$\epsilon_f$0.05-10dx | $6.025 \times 10^4$ | 31.70 | 0.01 | 0.05 | $1.43 \times 10^{-2}$ | 14 | 32.10 |
| 2E14-$\epsilon_r$0.1-$\epsilon_f$0.05-5dx | $6.025 \times 10^4$ | 31.70 | 0.1 | 0.05 | $1.43 \times 10^{-2}$ | 14 | 32.10 |
| 2E14-$\epsilon_r$0.1-$\epsilon_f$0.05-10dx | $6.025 \times 10^4$ | 31.70 | 0.1 | 0.05 | $1.43 \times 10^{-2}$ | 14 | 32.10 |
| 2E14-$\epsilon_r$0.01-$\epsilon_f$0.001-5dx | $6.025 \times 10^4$ | 31.70 | 0.01 | 0.001 | $1.43 \times 10^{-2}$ | 14 | 32.10 |
| 2E14-$\epsilon_r$0.01-$\epsilon_f$0.0001-5dx | $6.025 \times 10^4$ | 31.70 | 0.01 | 0.001 | $1.43 \times 10^{-2}$ | 14 | 32.10 |
| 3E15-no-feedback | $1.050 \times 10^3$ | 0.50 | None | None | $4.24 \times 10^{-3}$ | 15 | 0.80 |
| 3E15-$\epsilon_r$0.1-$\epsilon_f$0.05-5dx | $1.050 \times 10^3$ | 0.50 | 0.1 | 0.05 | $8.30 \times 10^{-3}$ | 15 | 0.80 |
| 3E15-$\epsilon_r$0.01-$\epsilon_f$0.05-5dx | $1.050 \times 10^3$ | 0.50 | 0.01 | 0.05 | $8.30 \times 10^{-3}$ | 15 | 0.80 |
| 3E15-$\epsilon_r$0.01-$\epsilon_f$0.001-5dx | $1.050 \times 10^3$ | 0.50 | 0.01 | 0.001 | $8.30 \times 10^{-3}$ | 15 | 0.80 |
| 3E16-no-feedback | $1.050 \times 10^3$ | 0.50 | None | None | $4.14 \times 10^{-3}$ | 16 | 0.80 |
| 3E16-$\epsilon_r$0.1-$\epsilon_f$0.05-5dx | $1.050 \times 10^3$ | 0.50 | 0.1 | 0.05 | $4.14 \times 10^{-3}$ | 16 | 0.80 |
| 3E16-$\epsilon_r$0.1-$\epsilon_f$0.05-5dx | $1.050 \times 10^3$ | 0.50 | 0.1 | 0.05 | $4.14 \times 10^{-3}$ | 16 | 0.80 |
| 3E17-$\epsilon_r$0.01-$\epsilon_f$0.001-5dx | $1.050 \times 10^3$ | 0.50 | 0.01 | 0.001 | $4.14 \times 10^{-3}$ | 16 | 0.80 |
| 3E17-no-feedback | $1.050 \times 10^3$ | 0.50 | None | None | $2.07 \times 10^{-3}$ | 17 | 0.80 |
| 3E17-$\epsilon_r$0.1-$\epsilon_f$0.05-5dx | $1.050 \times 10^3$ | 0.50 | 0.1 | 0.05 | $2.07 \times 10^{-3}$ | 17 | 0.80 |
| 3E17-$\epsilon_r$0.01-$\epsilon_f$0.05-5dx | $1.050 \times 10^3$ | 0.50 | 0.01 | 0.05 | $2.07 \times 10^{-3}$ | 17 | 0.80 |
| 3E17-$\epsilon_r$0.01-$\epsilon_f$0.001-5dx | $1.050 \times 10^3$ | 0.50 | 0.01 | 0.001 | $2.07 \times 10^{-3}$ | 17 | 0.80 |
| 4E16-no-feedback | $6.039 \times 10^3$ | 6.50 | None | None | $4.24 \times 10^{-3}$ | 16 | 7.00 |
| 4E16-$\epsilon_r$0.1-$\epsilon_f$0.05-5dx | $6.039 \times 10^3$ | 6.50 | 0.1 | 0.05 | $4.24 \times 10^{-3}$ | 16 | 7.00 |
| 4E16-$\epsilon_r$0.1-$\epsilon_f$0.05-7dx | $6.039 \times 10^3$ | 6.50 | 0.1 | 0.05 | $4.24 \times 10^{-3}$ | 16 | 7.00 |
| 4E16-$\epsilon_r$0.1-$\epsilon_f$0.05-10dx | $6.039 \times 10^3$ | 6.50 | 0.1 | 0.05 | $4.24 \times 10^{-3}$ | 16 | 7.00 |
| 4E16-$\epsilon_r$0.01-$\epsilon_f$0.05-5dx | $6.039 \times 10^3$ | 6.50 | 0.01 | 0.05 | $4.24 \times 10^{-3}$ | 16 | 7.00 |
| 4E16-$\epsilon_r$0.01-$\epsilon_f$0.01-5dx | $6.039 \times 10^3$ | 6.50 | 0.01 | 0.01 | $4.24 \times 10^{-3}$ | 16 | 7.00 |
| 4E16-$\epsilon_r$0.01-$\epsilon_f$0.001-5dx | $6.039 \times 10^3$ | 6.50 | 0.01 | 0.001 | $4.24 \times 10^{-3}$ | 16 | 7.00 |
| 4L16-$\epsilon_r$0.1-$\epsilon_f$0.05-5dx | $6.040 \times 10^3$ | 6.90 | 0.1 | 0.05 | $4.24 \times 10^{-3}$ | 16 | 7.00 |
| 4L16-$\epsilon_r$0.01-$\epsilon_f$0.05-5dx | $6.040 \times 10^3$ | 6.90 | 0.01 | 0.05 | $4.24 \times 10^{-3}$ | 16 | 7.00 |
| 4L16-$\epsilon_r$0.01-$\epsilon_f$0.001-5dx | $6.040 \times 10^3$ | 6.90 | 0.01 | 0.001 | $4.24 \times 10^{-3}$ | 16 | 7.00 |

Fig. 4 shows the phase evolution of the gas within 0.5 pc of the BH for four simulations from Accretion Event 2. The first plot shows the state of the gas before feedback is activated, while the other three show the impact of feedback 200 kyr later at progressively higher levels of feedback efficiency. Feedback heats the gas near the particle up to $10^8$ K, with an extended range of densities at $T \simeq 10^4$ K, likely due to the peak in the pristine gas cooling curve at $T = 2 \times 10^4$ K (Thoul & Weinberg 1994). Additionally, a cluster of more massive cells is visible in the range $10^7$ K $< T < 10^8$ K, where cooling starts to increase in efficiency again. A small portion of very hot, low-density cells is expected when feedback is included in cosmological simulations (Valentini et al. 2017; Oppenheimer et al. 2021).

The clumping factor, defined as $C = \langle \rho^2 \rangle / \langle \rho \rangle^2$, is shown for each feedback scenario as a measure of how concentrated the gas is relative to a uniform distribution. Simulations with maximum feedback exhibit a lower clumping factor ($C = 1.60$), indicating that feedback homogenizes the gas, while simulations with minimum feedback show higher clumping factors ($C = 2.05$), suggesting greater gas concentration.[1] Another clear impact of feedback is the reduction in high-density gas ($n > 1 \times 10^7$ cm$^{-3}$), with the mass halving from minimum to mid-range efficiencies, and further decreasing by 25 per cent from mid to maximum feedback levels. This demonstrates the self-regulating nature of thermal feedback, which allows the BH's gravitational sphere of influence to be periodically replenished while preventing overcooling and excessive clump formation. Though not shown, the no-feedback simulation at 31.90 Myr has the highest clumping factor at $C = 2.86$, closely followed by 2E14-$\epsilon_r$0.01-$\epsilon_f$0.05-7dx and 2E14-$\epsilon_r$0.01-$\epsilon_f$0.0001-5dx, where $C = 2.32$. Increasing the feedback injection radius from $r_{\text{fb}} = 5$ d$x$ to $r_{\text{fb}} = 7$ d$x$ means that less energy is being deposited per cell, hence the intensity of the feedback is curtailed. 2E14-$\epsilon_r$0.01-$\epsilon_f$0.05-7dx also has double the quantity of mass at high density compared to 2E14-$\epsilon_r$0.01-$\epsilon_f$0.05-5dx and a higher BH mass, though the shape of the phase diagram is similar, so gas is still being heated and blown out of the accretion region. The effectiveness of such feedback is discussed in Section 3.5.

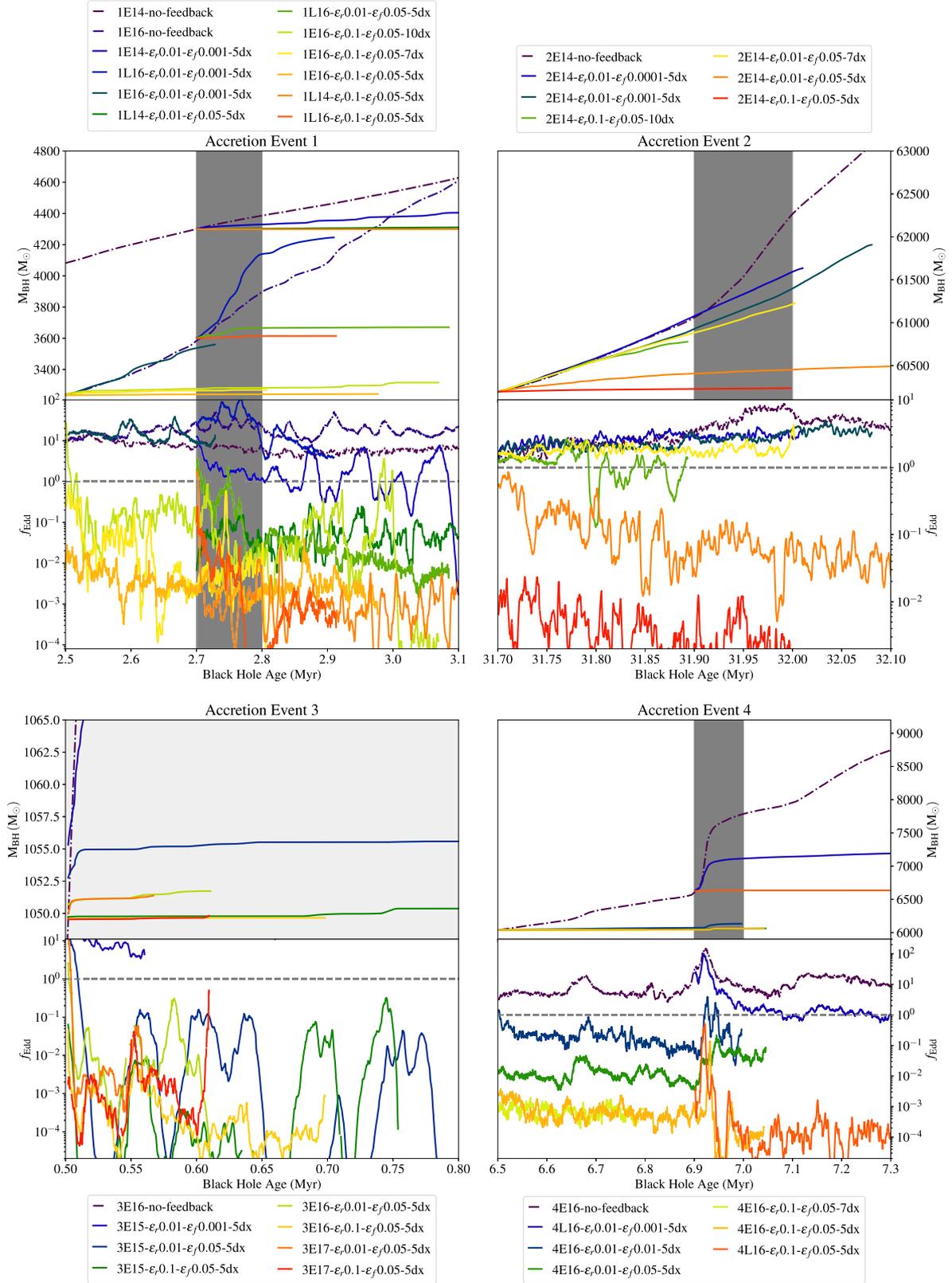

**Figure 1.** Time series of BH mass (top panels) and the accretion rate as a fraction of the Eddington rate (lower panels) for each accretion event. The dashed grey lines indicate $f_{\mathrm{Edd}} = 1$. The dark grey patches represent the extend of the accretion event, with Event 3 lasting the entire period shown. All data have been averaged over 100 data points or about 1000 yr.

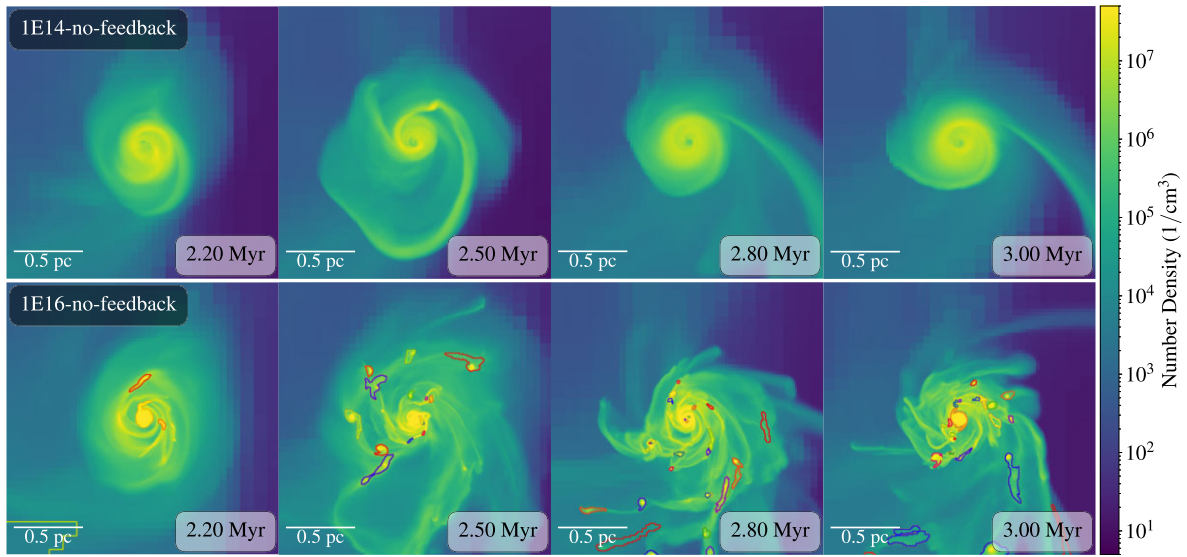

**Figure 2.** Time series of number density projections for 1E14-no-feedback (top row) and 1E16-no-feedback (bottom row) with clumps overlaid. The former has no distinct clumps and its BH grows through smooth accretion from the disc.

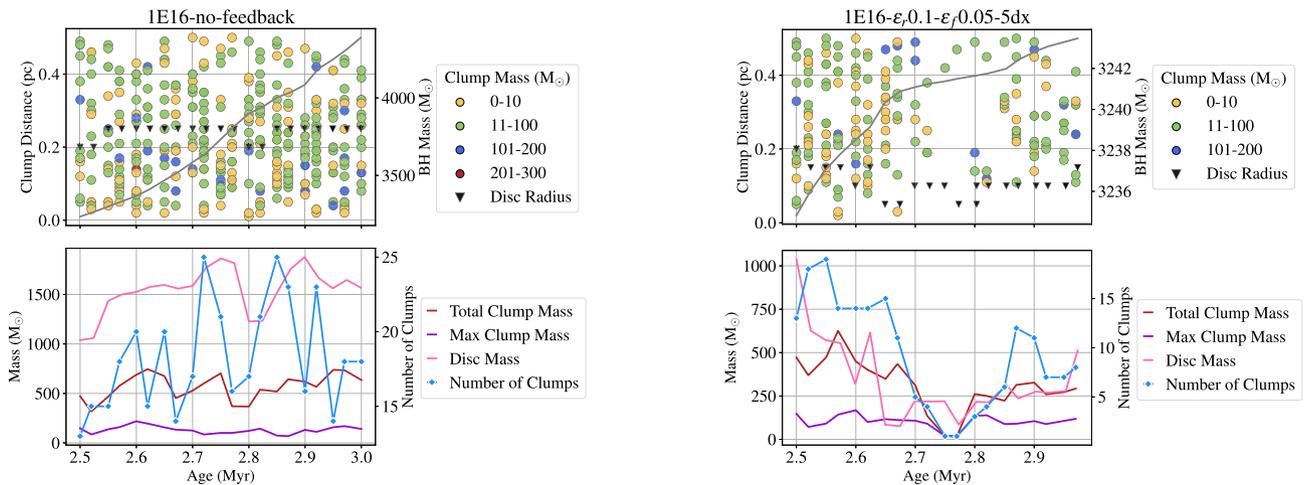

**Figure 3.** Clump evolution comparison between 1E16-no-feedback (left) and 1E16-$\epsilon_r$0.1-$\epsilon_f$0.05-5dx (right) during Accretion Event 1. The top panels show a scatter of clump distance from the BH particle as a function of age on the left *y*-axis. The colour of the scatter point indicates the mass bin to which the clump belongs. The disc radius is marked by solid black triangles. The right *y*-axis shows BH mass as a *grey solid line*. The bottom panels depict the total clump, max clump, and disc mass on the left *y*-axis and the number of clumps over time on the right *y*-axis. All data are gathered from within $r \leq 0.5$ pc of the BH particle.

### 3.4 Conditions for super-Eddington accretion

We find that super-Eddington growth is possible in a few scenarios. First, a very weak thermal feedback efficiency of $\epsilon_r = 0.01$ and $\epsilon_f = 0.001$ in *almost* all environments with the fiducial injection radius of $r_{fb} = 5dx$ results in super-Eddington accretion. The only exception is 1L14-$\epsilon_r$0.01-$\epsilon_f$0.001-5dx, which accretes at a trans-Eddington rate on average. In contrast, its higher resolution counterpart 1L16-$\epsilon_r$0.01-$\epsilon_f$0.001-5dx accretes over 10 times the Eddington rate. The former is more sensitive to feedback due to its lack of disc fragmentation, as explained previously in Section 3.2. All other simulations studied contain cool, dense clumps with high optical depths that act as absorbers of thermal energy, diminishing its effectiveness in heating and dispersing the surrounding gas.

---

[1]While clumping factors have a high variance over time (up to 25 per cent from the mean over 0.3 Myr), the ranking of simulations from lowest to highest stated does not change when averaged over the period.

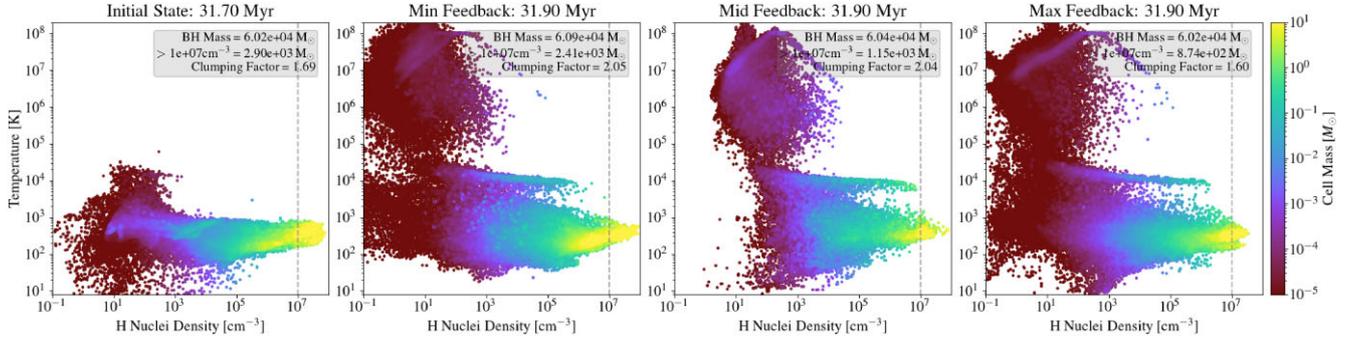

**Figure 4.** Phase plots of the gas within a 0.5 pc radius of the BH for Accretion Event 2. From left to right, the simulations are 2E14-no-feedback, 2E14-$\epsilon_r$0.01-$\epsilon_f$0.001-5dx, 2E14-$\epsilon_r$0.01-$\epsilon_f$0.05-5dx, and 2E14-$\epsilon_r$0.1-$\epsilon_f$0.05-5dx. Feedback heats gas near the particle up to $10^8$ K. There is an extended range of densities at $T \simeq 10^4$ K, which may be due to the peak in the pristine gas cooling curve at $T = 2 \times 10^4$ K (Thoul & Weinberg 1994). There is also a cluster of more massive cells in the range $10^7$ K $< T < 10^8$ K, where cooling starts to rise in efficiency once again.

Moreover, super-Eddington accretion is seen in medium-to-high thermal efficiencies for the $6 \times 10^4$ M$_\odot$ BH in Accretion Event 2 with $r_{fb} = 7dx/10dx$. Fig. 5 depicts projections of the temperature and density of the disc for three sets of simulations with the same feedback efficiency $\epsilon_r = 0.01$ and $\epsilon_f = 0.05$ (mid-range), but different injection radii; $r_{fb} = 5dx, 7dx/10dx$. Note that 2E14-$\epsilon_r$0.01-$\epsilon_f$0.05-5dx averages at a trans-Eddington accretion rate, while the last two average at super-Eddington. In the first panel, we can see that the pc-scale environment has been impacted by feedback; shocks and dissipation of the disc are visible, some regions have been heated to $T > 1 \times 10^5$ K, and a bubble of sparse, warm gas has formed around the BH, which previously had the densest gas concentration. However, the zoomed-in final column reveals that a dense, cold sub-pc inner disc (which is orthogonal to the outer disc) remains intact. This structure is robust enough to withstand thermal pressure, even in 2E14-$\epsilon_r$0.1-$\epsilon_f$0.05-5dx (not shown), where the feedback is 10 times more intense. This helps prevent the Eddington fraction from dropping sharply to $f_{Edd} < 10^{-3}$ once feedback is activated, as in Accretion Events 1 and 3 for the same level of feedback efficiency. The key takeaway is that even under significant feedback, the built-up environment of the MBH (usually defined in the range $\simeq 10^4$–$10^6$ M$_\odot$) is able to stabilize its accretion flow by maintaining a cold, dense core. The differences in feedback radii may influence the large-scale dynamics, but the inner disc's integrity is central to sustaining accretion in all feedback configurations.

The Accretion Event 4 simulation 4E16-$\epsilon_r$0.01-$\epsilon_f$0.01-5dx also achieves super-Eddington accretion at an average of $f_{Edd} = 1.58$ over 0.56 Myr of evolution. This $6 \times 10^3$ M$_\odot$ BH does not have such a built-up environment as for the $6 \times 10^4$ M$_\odot$ BH in Accretion Event 1, but the dense clumps present in this higher resolution run (dx = $4.24 \times 10^{-3}$ pc) provide some shielding from thermal feedback (as discussed in Section 3.2). This is the only simulation to achieve super-Eddington rates with $r_{fb} = 5dx$. The significant accretion event seen in the no-feedback iteration is also felt with thermal feedback present, briefly driving the Eddington ratio up to 5 and causing a considerable bump in the BH mass.

Trans-Eddington growth is possible for the $3 \times 10^3$ M$_\odot$ BH of Accretion Event 1, which is nestled within a disc of comparable mass (see Fig. 2 for projections). With $r_{fb} = 5dx$, $\epsilon_r = 0.01$, and $\epsilon_f = 0.05$ in 1L16-$\epsilon_r$0.01-$\epsilon_f$0.05-5dx, an average Eddington fraction $f_{Edd} = 1 \times 10^{-1}$ is maintained over 0.20 Myr. With $r_{fb} = 10dx$, $\epsilon_r = 0.1$, and $\epsilon_f = 0.05$ in 1L16-$\epsilon_r$0.1-$\epsilon_f$0.05-10dx, an average Eddington fraction $f_{Edd} = 0.2$ is maintained over 0.32 Myr. Trans-Eddington accretion is also maintained over 0.49 Myr for the $6 \times 10^3$ M$_\odot$ BH in the Accretion Event 4 simulation 4E16-$\epsilon_r$0.01-$\epsilon_f$0.05-5dx. This and 2E14-$\epsilon_r$0.01-$\epsilon_f$0.05-5dx are the only two mid-efficiency feedback trans-Eddington simulations with $r_{fb} = 5dx$. The effectiveness of thermal feedback to alter the state of the gas, drive outflows, and suppress accretion at lower efficiencies and larger injection radii will be discussed in Section 3.5.

Fig. 6 shows the fraction of time spent in each accretion phase (sub-Eddington, trans-Eddington, and super-Eddington) along with the fraction of mass gained over the total time in each mode for a sample of the super- and trans-Eddington accreting BHs. In the two Event 1 simulations, where the BHs spend most of their time in the super-Eddington regime $f_{Edd} > 1$, the majority of mass gain occurs during this phase, which is expected. These $\sim 100$ kyr periods of sustained supercritical growth illustrate that thermal feedback may not be strong enough to quench further accretion soon after. It is worth remarking that for two of the three BHs that average in the trans-Eddington phase (where $0.01 < f_{Edd} < 1$), the super-Eddington mode still accounts for the majority of the mass gain, despite the BH spending more time in lower accretion states. This suggests that brief periods of super-Eddington accretion are disproportionately important for BH growth. Even spending as little as 5 per cent of the time in $f_{Edd} > 1$ is sufficient to contribute 50 per cent of the mass gain, as observed in the simulation 1L16-$\epsilon_r$0.1-$\epsilon_f$0.05-5dx. However, longer term evolution past the accretion event could balance contributions from different phases [as suggested by the results from 1E16-$\epsilon_r$0.1-$\epsilon_f$0.05-5dx (0.48 Myr)]. These results reinforce the idea that even short periods of super-Eddington accretion are an extremely efficient vehicle for BH growth.

### 3.5 The effectiveness of thermal feedback

What does effective thermal feedback look like? The non-linear processes governing accretion physics make it difficult to deduce the average feedback efficiency of the first generation of BHs *ab*

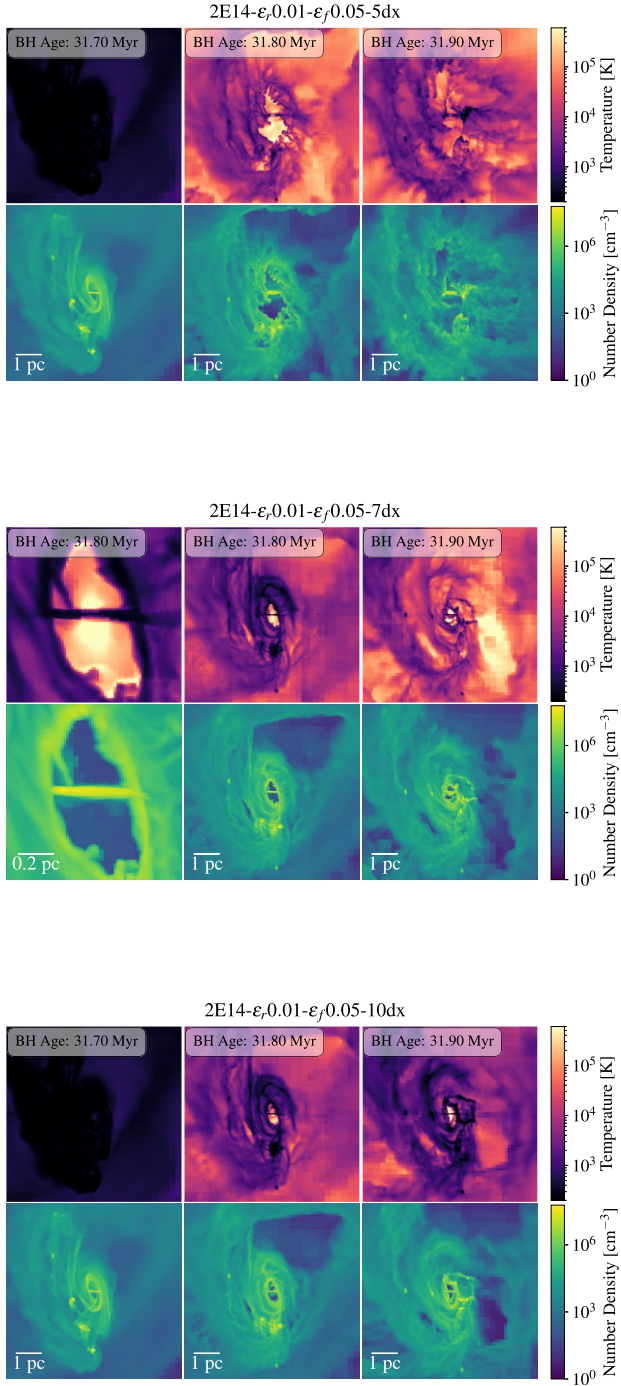

**Figure 5.** Time series projections of density and temperature for Event 2 across simulations with increasing feedback injection radii but the same feedback efficiencies. The top panel represents the fiducial feedback injection radius of $r_{fb} = 5\,dx$, followed by $r_{fb} = 7\,dx$ in the middle panel, and $r_{fb} = 10\,dx$ in the bottom panel. In all simulations, feedback heats and ejects gas near the BH. However, as the feedback energy is spread over a larger volume, its impact on the gas and disc diminishes. This leads to a structure that becomes warmer but remains intact, as shown in the final row. Notably, a compact inner sub-pc disc, visible in the first column of the second row, persists across all three simulations.

*initio*. Numerical simulations often calibrate feedback efficiencies to produce BHs with properties that align with observed BH–galaxy scaling relations at low redshift (e.g. $M_{BH}$–$\sigma$). However, many of these relations may not hold true for the very high redshift $z > 10$ proto-galactic environments in this work. In lacking observable benchmarks, we assess the ability of feedback to heat and regulate the gas dynamics by coupling energy to the gas, preventing rapid cooling and driving outflows by comparing the simulations to their no-feedback counterparts and each other.

Fig. 7 shows the cumulative mass enclosed around the BH as a function of radius for four Accretion Event 2 simulations: the initial state at 31.70 Myr with a BH mass of 60, 205 $M_\odot$ for reference, and three feedback scenarios of increasing intensity. The mass gained by each simulation over 200 kyr is written in brackets. The mass enclosed rises sharply up to ~1.2 pc and plateaus around $10^4\,M_\odot$, which roughly corresponds to the radius of the disc (see Fig. 5). As feedback intensity increases, the total enclosed mass decreases, indicating more effective gas expulsion at higher efficiencies. The lowest feedback run (blue line) shows the highest mass retention, hardly deviating from the initial state, which suggests that the feedback is not pushing out much gas, if at all. In contrast, the mid-range (purple) and maximum (crimson) feedback runs have up to two orders of magnitude less mass at the lowest radii, and about half the mass enclosed within the disc at 0.2 pc – indicating that stronger feedback effectively suppresses mass accumulation, particularly within the disc. While the overall trend suggests that the highest feedback intensity should expel more gas, there are a few points within 0.2 pc where moderate feedback results in slightly less mass enclosed. Moreover, the hump at 0.2 pc indicates that moderate feedback results in a slightly less aggressive interaction with the disc, preserving more gas within its gravitational bounds. These subtle differences reflect the complexity of gas dynamics in feedback-regulated environments. None the less, a clear delineation emerges between the minimum and moderate/maximum feedback ranges for a given injection radius, suggesting that the former is ineffective at driving outflows.

Fig. 8 shows the ratio of densities enclosed to the initial densities enclosed before feedback was activated, versus the feedback efficiency normalized by the physical volume of the feedback region. The values are given ~0.1 Myr after feedback is activated. This 'feedback intensity' is defined as $\epsilon_r\epsilon_f/V_{fb}$, where the denominator is the volume of the feedback sphere in units of pc$^3$. Lower values might indicate longer distance but milder effects, where feedback impacts the environment over a wider area but with less intensity, and higher values imply that a significant amount of feedback is concentrated in a relatively small spatial region. The radius within which the mass enclosed per event is measured is defined as half of the radius at which the mass enclosed converges with the initial state; $r = 0.2, 0.6, 0.2$ pc for accretion events 1, 2, and 4, respectively, which we determine with plots such as Fig. 7. Each point represents a simulation, which is labelled with the event number, the main feedback parameter from the fiducial setting of $\epsilon_r = 0.01$ and $\epsilon_f = 0.05$, and the radius of the feedback sphere in units of cell widths. The points are coloured according to the relative density of their feedback region in comparison to the density just before feedback is activated. The overall y-weighted least-squares linear regression trendline shows a decline in relative mass enclosed as the feedback efficiency per unit volume increases. This suggests a correlation between high feedback efficiency and the ability to drive outflows away from the BH. The relative density of the feedback sphere also supports this; regions with density ratios lower than $\leq 10^{-1}$ correspond with high feedback intensity. The considerable scatter around the trend reflects variations in local conditions; for instance, the 'L' simulations, which are initiated <0.1 Myr before the accretion event, generally enclose less mass relative to the later starting position than those that began ≥0.2 Myr prior, even though all measurements are made $\simeq$0.15–0.17 Myr post feedback activation. The higher accretion rate close to the event drives stronger feedback that may evacuate gas from the region faster in comparison to a more quiescent period.

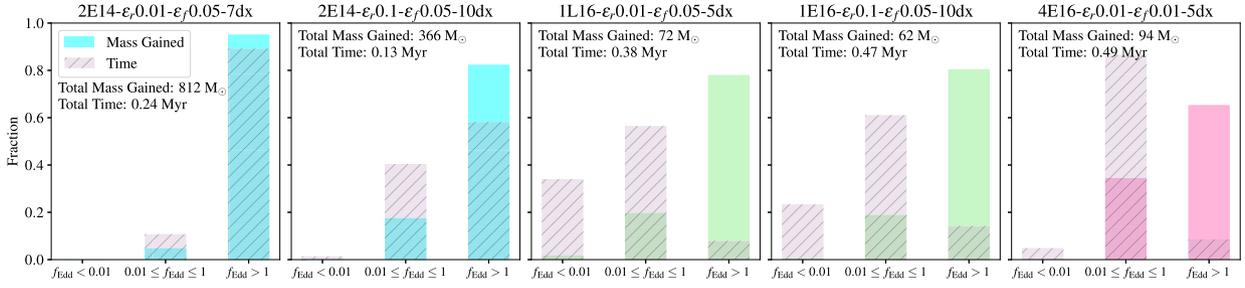

**Figure 6.** Fractional values of mass gained (*solid colour*) and time spent (*hatched*) in one of three accretion modes: sub-Eddington ($f_{\rm Edd} < 0.01$), trans-Eddington ($0.01 \leq f_{\rm Edd} \leq 1$), and super-Eddington ($f_{\rm Edd} > 1$). The duration of the accretion period is indicated on each panel as 'Total Time' along with the 'Total Mass Gained'. Despite some simulations spending little time in the super-Eddington regime, this mode dominates the mass growth in most scenarios.

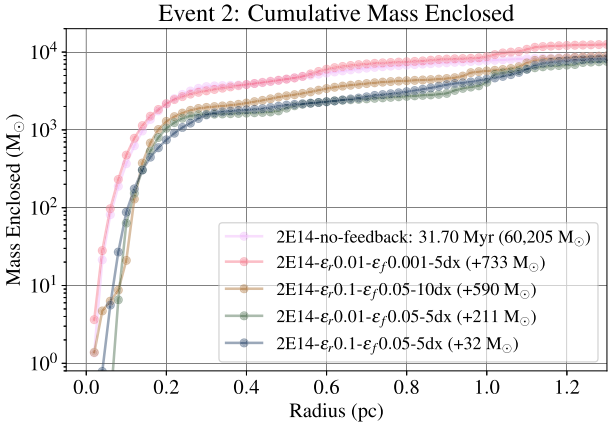

**Figure 7.** The cumulative mass enclosed around BHs in Event 2 simulations under various thermal feedback scenarios. The initial state at 31.70 Myr (pale pink) serves as a reference, with the BH mass at 60,205 $M_\odot$. The other four simulations are measured at 31.90 Myr, after 200 kyr of evolution in the presence of feedback. The value in brackets indicates the mass gained over this period. Minimum feedback efficiency (coral) results in the highest mass retention, while the two maximum feedback efficiency runs (green and navy) are the most effective at reducing the enclosed mass. The mid-range configuration (mustard) exhibits intermediate behaviour. The cumulative mass enclosed decreases with increasing feedback efficiency, indicating more effective gas expulsion at higher efficiencies.

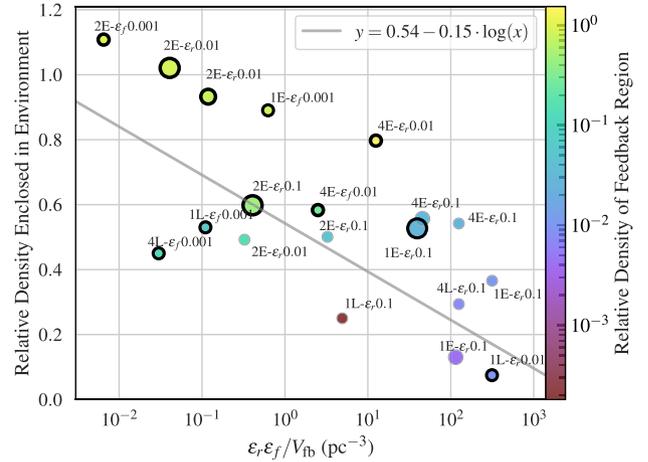

**Figure 8.** The ratio of density enclosed by the current simulation to the initial density enclosed before feedback was activated [different for the 'early' (E) and 'late' (L) simulations] versus the feedback efficiency normalized by the physical volume of the feedback region after 0.15–0.17 Myr of evolution. The radius within which the mass enclosed per event is measured is defined as half of the radius at which the mass enclosed converges with the initial state; $r = 0.2, 0.6, 0.2$ pc for accretion events 1, 2, and 4, respectively. The scatter points are coloured by the ratio of the density of the feedback region to the original density before feedback. The thick black circled points indicate those that accrete at trans- or super-Eddington rates on average. This trendline can be used to help determine which simulation feedback parameters lead to effective thermal feedback; points on the top left and in light colours are the least effective while points on the bottom right and in dark colours are the most effective.

We focus our discussion on identifying the subset of black circled points in Fig. 8 that accrete at trans- or super-Eddington rates and also have effective thermal feedback. We draw from previous analyses in this work to reach the simulation classifications. First, the point labelled '2E-$\epsilon_f$0.001' at (0.8, 0.9) corresponds to 1E16-$\epsilon_r$0.01-$\epsilon_f$0.001-5dx and averages at an $f_{\rm Edd} = 19$ accretion rate. However, from its position in the top left corner and the high relative density of its feedback region, the thermal feedback seems too weak to convert to kinetic energy and affect change in the BH's environment. This applies to the other yellow points nearby too (2E14-$\epsilon_r$0.01-$\epsilon_f$0.01-7dx and 2E14-$\epsilon_r$0.01-$\epsilon_f$0.01-10dx), for which you can see temperature and density projections in Fig. 5. With little change in the density of the feedback region, 4E16-$\epsilon_r$0.01-$\epsilon_f$0.05-5dx at (15, 0.8) also has ineffective feedback, despite it having moderate feedback efficiency. This sets it apart from the trend, though some scatter is to be expected when quantifying small-scale gas dynamics at high resolution. On the other end of the relation, 1L16-$\epsilon_r$0.01-$\epsilon_f$0.05-5dx is located at (314, 0.07) in the bottom right quadrant and has a feedback region relative density of about 0.01. This simulation shows strong evidence of effective thermal feedback while maintaining a trans-Eddington accretion rate throughout its 2.93 Myr of evolution (see middle panel of Fig. 6). Though it does not push out as much gas in comparison, 1E16-$\epsilon_r$0.1-$\epsilon_f$0.05-10dx, which is located at (70, 0.55), is also in the bottom right quadrant and has a feedback region relative density of $\simeq$0.03, having pushed out 97 per cent of the gas. It maintains a slightly lower Eddington ratio of $f_{\rm Edd} = 0.13$. We categorize both of these simulations as having effective thermal feedback.

**Table 3.** Subset of simulations that accrete at trans- or super-Eddington rates on average. The columns are as follows: simulation name, average Eddington accretion ratio, duration of simulation, and whether or not the thermal feedback is effective based on the analysis in Section 3.5. Note that there are more simulations with $\epsilon_r = 0.01$ and $\epsilon_f = 0.001$ that accrete at these rates, but we include only those with the maximum and minimum $f_{Edd}$ for brevity; we categorize all such simulations as having ineffective thermal feedback.

| Simulation | $f_{Edd}$ | Time (Myr) | Effective feedback? |
|---|---|---|---|
| 1E16-$\epsilon_r$0.01-$\epsilon_f$0.001-5dx | 19.0 | 0.23 | No |
| 1L16-$\epsilon_r$0.01-$\epsilon_f$0.001-5dx | 33.5 | 0.21 | No |
| 1L16-$\epsilon_r$0.01-$\epsilon_f$0.05-5dx | 0.32 | 0.38 | Yes |
| 1E16-$\epsilon_r$0.1-$\epsilon_f$0.05-10dx | 0.13 | 0.47 | Yes |
| 2E14-$\epsilon_r$0.01-$\epsilon_f$0.05-5dx | 0.43 | 0.84 | Yes |
| 2E14-$\epsilon_r$0.01-$\epsilon_f$0.05-7dx | 2.66 | 0.24 | No |
| 2E14-$\epsilon_r$0.01-$\epsilon_f$0.05-10dx | 2.60 | 0.48 | No |
| 2E14-$\epsilon_r$0.1-$\epsilon_f$0.05-10dx | 2.09 | 0.13 | Maybe |
| 4E16-$\epsilon_r$0.01-$\epsilon_f$0.05-5dx | 0.32 | 0.49 | No |
| 4E16-$\epsilon_r$0.01-$\epsilon_f$0.01-5dx | 1.58 | 0.56 | Maybe |

Clustered in the middle of Fig. 8 are four Eddington-accreting simulations that eject between 40 per cent and 55 per cent of the gas in the disc, but keep 15–45 per cent of the gas in the feedback region. Is this sufficiently disruptive to be considered effective feedback? We must look to other analyses to judge this. From the first row of Fig. 5, we can see that 2E14-$\epsilon_r$0.01-$\epsilon_f$0.05-5dx does appear to drive outflows on pc scales, destroying most of the outer disc. Fig. 7 shows that it has a very similar enclosed mass profile to its higher efficiency counterpart, 2E14-$\epsilon_r$0.1-$\epsilon_f$0.05-5dx. These two pieces of evidence tip the balance in favour of effective feedback. Conducting a similar study on 2E14-$\epsilon_r$0.1-$\epsilon_f$0.05-10dx reveals that it too has a similar projection plot and mass enclosed to 2E14-$\epsilon_r$0.1-$\epsilon_f$0.05-5dx. On the other hand, 4E16-$\epsilon_r$0.01-$\epsilon_f$0.01-5dx has made little impact on its host disc in terms of disrupting its size and dynamics, though the cumulative mass enclosed profile shows that it has reduced the mass within 0.65 pc of the BH by the same amount as the maximum efficiency runs. However, it encloses more mass within 1.5 pc, implying that it has not been as effective at expelling gas on pc scales. These considerations tip the balance against 4E16-$\epsilon_r$0.01-$\epsilon_f$0.01-5dx having effective feedback. Likewise, the cumulative mass enclosed profile of 1L16-$\epsilon_r$0.01-$\epsilon_f$0.001-5dx shows that it has hardly expelled more gas than the no-feedback run. See Appendix A for the cumulative mass profiles and projections of 4E16-$\epsilon_r$0.01-$\epsilon_f$0.01-5dx and 1L16-$\epsilon_r$0.01-$\epsilon_f$0.001-5dx. We summarize our classification of all trans- and super-Eddington accreting BHs in Table 3. Notably, if 4E16-$\epsilon_r$0.01-$\epsilon_f$0.01-5dx continues to grow at its current average rate, it could explain some of the recent discoveries of $\simeq 10^8\,M_\odot$ SMBHs at $z \simeq 10$.

## 4 DISCUSSION

We have shown that super-Eddington accretion is possible for BHs in the range $10^3\,M_\odot < M_{BH} < 10^5\,M_\odot$ on short time-scales <1 Myr in an early Universe environment, but can it be sustained? Fig. 9 displays the BH mass growth across cosmic time for each simulation that accreted at trans- or super-Eddington rates and had 'Yes' or 'Maybe' classifications for effective thermal feedback, assuming that they continue to grow at their current averaged rates. If this were possible, they could explain a subset of observations (all other data points in Fig. 9) from Harikane et al. (2023), Maiolino et al. (2024a), and Greene et al. (2024). We now discuss other works that have focused on longer term efficient accretion at lower spatial resolutions to evaluate the likelihood of this scenario.

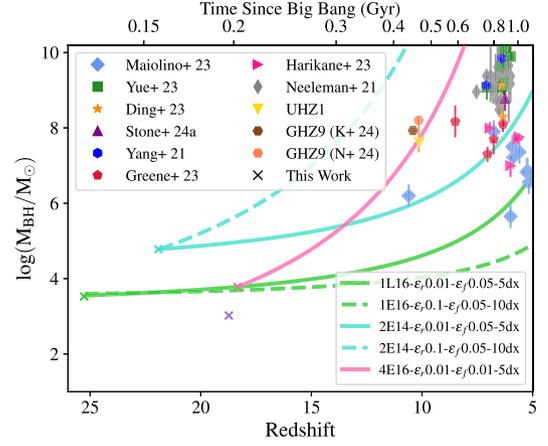

**Figure 9.** The base 10 logarithm of BH mass versus redshift (lower *x*-axis) and time in Gyr (upper *x*-axis). The crosses indicate the starting point of the four accretion events studied in this work, defined from no-feedback runs: 1 (*green*), 2 (*cyan*), 3 (*lilac*), and 4 (*pink*). The other marked points represent observational data from surveys : Neeleman et al. (2021), Yang et al. (2021), Ding et al. (2023), Harikane et al. (2023), Yue et al. (2023), Maiolino et al. (2024a), Greene et al. (2024), and Stone et al. (2024).Two recent exciting finds are highlighted: UHZ1 (Bogdán et al. 2024); GHZ9 (K+ 24) (Kovács et al. 2024) and GHZ9 (N+ 24) (Napolitano et al. 2024). The lines emanating from three of the crosses are each of the 'Yes' or 'Maybe' simulations in Table 3, assuming that their current average accretion rate continues for millions of years. If this were possible, they could explain a subset of observations from Harikane et al. (2023), Maiolino et al. (2024a), Greene et al. (2024), and even the some of earliest BH candidates detected to date, UHZ1 and GHZ9.

Mehta, Regan & Prole (2024) demonstrate that super-Eddington accretion on to stellar-mass BHs in a compact, high-z galaxy can occur over millions of years with supernova feedback (a single momentum injection) included – provided the BHs are situated in gas-rich environments. They find that the radially inward flow of the cool, dense gas exceeds the outward thrust of the supernova explosion, and the mixing of gas flows conspires to reduce angular momentum. In a minority of cases, nearby supernova shocks enhance growth, creating inhomogeneities around the BH particle. Their gas element resolution of $dx \simeq 1 \times 10^{-2}$ pc is comparable to the lowest resolution simulations in this work, for which we found thermal feedback to have a greater adverse effect on growth than at higher resolutions due to the less dense clumps. While we have not included supernova feedback, these findings indicate that it may not significantly stifle super-Eddington growth as the seed evolves, at least on time-scales of $\sim 10^6$ Myr. It should be noted that the authors sum the BHL formula and a term accounting for turbulence to calculate the overall rate, leading to a greater estimation of accretion on to the BH than used in this work.

In a similar vein, Lupi et al. (2016) show that stellar-mass BHs (20 or 100 M$_\odot$) bound to clumps in the circumnuclear disc of a more advanced galaxy experience periodic super-Eddington growth over 3 Myr, even with thermal feedback present. Once again, the gas resolution is comparable to the lowest resolution simulations present in this work, $dx \simeq 10^{-1/-2}$ pc. They attribute this rapid growth to the low radiative efficiency of supercritical flows as well as gas-rich environs. They compute the radiative efficiency $\epsilon_r$ on the fly based on the Eddington ratio, with high ratios corresponding to low $\epsilon_r$. This results in frequent drops to $\epsilon_r \ll 0.01$. With a fixed $\epsilon_f = 0.15$, the overall efficiency of feedback is $\epsilon_f \epsilon_r \ll 1.5 \times 10^{-3}$, which is less than our maximum feedback value. Their environment is most comparable to our Accretion Event 2, in which the MBH remains nestled within a compact inner disc, regardless of feedback efficiency. Our BH also experiences super-Eddington accretion at similar feedback efficiencies; however, $f_{\rm Edd} = 1$–$5$ in this work, in contrast to the $f_{\rm Edd} = 100$ reached in Lupi et al. (2016). This could be due to their on-the-fly feedback implementation, which enables more prolonged periods of super-Eddington growth and/or the presence of metal cooling in their lower redshift galaxy host. This work supports our findings that the BH being enveloped in dense gas significantly reduces the deleterious impact of thermal feedback on growth.

Lupi et al. (2024b) use cosmological zoom-in simulations at lower resolution (the Bondi radius is unresolved) to model a quasar at $z \sim 7$ including both radiative and kinetic feedback. Their results show that super-Eddington accretion can indeed be sustained for $\sim 10$ Myr, allowing $10^5 \, M_\odot$ MBHs to grow rapidly by up to three orders of magnitude. Again, lower feedback efficiency during super-Eddington phases makes this possible. A semi-analytical calculation shows that the MBH spin remains moderate during super-Eddington phases, which reduces feedback efficiency. They also find that a weaker magnetic field in the disc results in more efficient accretion – in particular, extended phases of super-Eddington growth – in comparison to a magnetically arrested disc. Since we are simulating the first generation of BHs born into pristine gas, our simulations are metal-free. The discs hosting the BHs are also less massive than a $z \sim 7$ quasar disc. This environment is less likely to support the formation of powerful, relativistic jets; hence, we have not included the kinetic feedback that impedes quasar growth so severely. On the other hand, metal line emission available to the quasar enables more efficient cooling and fragmentation of the gas. In essence, they were exploring the possibility of super-Eddington accretion in the unfavourable conditions of magnetically sourced jets, radiative feedback, and relatively high BH spin, whereas our unfavourable conditions are thermal feedback, lack of pressure from inflowing gas from larger scales (due to higher redshift), and lack of metal cooling. They find that the gas-rich conditions and short free-fall time allowed the accretion region to be replenished quickly post-jet feedback episodes, maintaining a density of $n \simeq 10^4$–$10^5 \, \rm cm^{-3}$ on average. However, they caution that their limited resolution may not properly resolve the shocked gas and so may be under-representing the impact of the feedback. Since our simulations resolve the Bondi radius by a factor of a few most of the time (see appendix A of G24), and we simulate in sub-pc gas resolution, we should be capturing shocks quite well. In simulations with strong thermal feedback, the accretion region can retain an average density of $n = 10^5 \, \rm cm^{-3}$ (1E16-$\epsilon_r$0.1-$\epsilon_f$0.05-5dx and 2E14-$\epsilon_r$0.1-$\epsilon_f$0.05-5dx), or it can quickly drop to $n < 10 \, \rm cm^{-3}$ (3L16-$\epsilon_r$0.1-$\epsilon_f$0.05-5dx). However, parts of the gas are heated to temperatures $> 10^8$ K (see Fig. 4).

Massonneau et al. (2023) explore the effect of different modes of feedback on a $z \simeq 4$ quasar of initial mass $10^6 \, M_\odot$ within an isolated galaxy simulation. In simulations where only thermal feedback is activated if $0.01 < f_{\rm Edd} < 1$, the BH grows rapidly during super-Eddington episodes over $\sim 10$ Myr. Without kinetic feedback, cold gas inflow is not efficiently counteracted, leading to sustained accretion above the Eddington limit. The BH triples its mass early on and grows $\simeq 15$ per cent more massive than in other simulations with kinetic feedback in this regime. While their total feedback efficiency $\epsilon_r \epsilon_f = 0.015$ is over double our maximum (0.005), their built up-environment is even more conducive to growth than in Lupi et al. (2024b). This highlights the significance of the gas composition when determining the effectiveness of feedback. Their kpc-scale disc and more efficient metal cooling can withstand and counteract the detrimental effects of thermal feedback on BH growth far more effectively than the sparse, pristine landscape of the early Universe mini-haloes, which are gas-poor and small in extent (Correa-Magnus et al. 2023). As the young BHs studied here grow and their host mini-haloes develop into galaxies, higher feedback efficiencies may still allow for super-Eddington growth.

To summarize, it has been shown that super-Eddington accretion can be sustained for $\simeq 3$ Myr in compact gas-rich environments (Lupi et al. 2016; Mehta et al. 2024) and on the scale of $\simeq 10$ Myr in lower resolution simulations (Massonneau et al. 2023; Lupi et al. 2024b), when feedback efficiencies can be suppressed and accreted gas is replenished frequently. Simulations incorporating spin (Lupi et al. 2024b) show that moderate BH spins during super-Eddington phases reduce feedback efficiency, further supporting extended growth. Our higher resolution simulations resolve the Bondi radius and the sub-pc gas structure, capturing shocks more effectively, but operate in pristine, high-redshift environments where gas-poor conditions limit sustained accretion. Collectively, these findings show promise for the simulations in Fig. 9 to maintain efficient growth across cosmic time-scales and potentially seed the SMBHs observed at $z \simeq 6$–$10$.

## 5 CAVEATS

The gas content of the mini-haloes in this work is likely too cool and abundant. In galaxy formation simulations, thermal feedback from supernovae is typically used to heat the interstellar medium, drive outflows, and regulate star formation. The effectiveness of this feedback is often evaluated by comparing simulated star formation rates with observed or expected values; rates exceeding expectations indicate ineffective feedback. However, this metric is not applicable to the early Universe environment explored in this study. To simplify the simulations, we have excluded both background feedback and star particle formation. As a result, the gas content represents an upper bound, as star particle formation would ordinarily remove gas from cells and assign it to particles to conserve mass-energy. However, star formation would necessitate increasing the resolution to sub-au scales to properly resolve small-scale fragmentation, which would considerably increase the computational expense and limit the evolution time achievable (Prole et al. 2021; Chon et al. 2024).

Furthermore, cooling from deuterium hydride (HD) is not included in the chemistry model. While HD dominates over $H_2$ at $T < 100$ K, collisional ionization in primordial gas is required to form the molecule in high abundance. This can occur in shock heating to temperatures above $10^4$ K, as in the virialization of atomic cooling haloes, or through the radiation from nearby Pop III stars (Greif et al. 2010) – or, in our case, thermal BH feedback. The BHs in this work form from 'the first star in the Universe' [a so-called Pop III.I star (Tan & McKee 2008)] in low-mass mini-haloes in which the HD fraction is not high enough to cool the gas to $\sim T_{\rm CMB}$ (Yoshida et al. 2007). However, thermal feedback heats and sometimes shocks the gas, which could generate enough HD to become the dominant cooling channel at low temperatures. It may be worth adding this species to the chemistry model in future work.

We have only modelled one type of BH feedback. While thermal feedback primarily heats and expels gas in the immediate vicinity of the BH, radiative feedback extends its influence much further, especially in the early universe. Electromagnetic radiation from the accretion disc, particularly in the ultraviolet and X-ray bands, not only heats and ionizes nearby gas but also exerts radiation pressure, more effectively clearing the region around the BH. Additionally, radiative feedback impacts larger scales by photoionizing distant gas in the halo, inhibiting its cooling and collapse, which can have significant consequences for the evolution of the surrounding environment. Furthermore, kinetic feedback in the form of jets can help transfer energy from small to large scales (Maiolino et al. 2012). Generally, high-velocity jets are observed to arise from early-type galaxies hosting MBHs with low accretion rates relative to their Eddington rate, while lower velocity outflows typically arise in systems with higher $f_{Edd}$ (Best & Heckman 2012; Heckman & Best 2014). However, Regan et al. (2019) find that BHs formed from SMSs initially undergo super-Eddington accretion. Bipolar jets triggered by this accretion evacuate gas within $\simeq 0.1$ pc of the BH, reducing the effective accretion rate to $f_{Edd} = 0.1$–$0.5$, similar to the effects of thermal feedback on our high-resolution simulations with thermal feedback.

The BHs in this work originate from Pop III stars that lived between 2 and 3.8 Myr prior to collapsing into BHs, though we do not simulate their main-sequence lifetime. Ionizing radiation has been shown to efficiently evacuate gas from the centre of the mini-halo, either dispersing it to the outskirts or destroying the gas structure completely (Shapiro, Iliev & Raga 2004; Alvarez, Bromm & Shapiro 2006; Whalen et al. 2008; Latif et al. 2022). This would have significantly reduced the density of the gas in the vicinity of the particles, potentially limiting their initial growth to a small fraction of the Eddington rate (Alvarez, Wise & Abel 2009). However, Jaura et al. (2022) find that the ionizing radiation from Pop III stars does not escape the dense accretion disc surrounding them when photons are injected into the simulation on scales smaller than the local scale height of the disc. This would imply that the inclusion of radiative feedback has little impact on the total mass of protostars formed during the 20 kyr simulated. Therefore, the nature of the environment in which the first generation of stellar-collapse BHs are born – whether gas-rich or gas-poor – remains uncertain.

While the BH masses at the beginning of the accretion periods are not quite MBHs (apart from Event 2), they are generally still too high to have originated from a standard collapse of a single Pop III stellar core. Our set-up of seeding a $270\,M_\odot$ BH, letting it evolve without feedback, and then resimulating with feedback complicates defining the assumed formation mechanism. Immediately introducing even weak feedback to the $270\,M_\odot$ direct-collapse seed quickly quelled accretion and we did not have the computational resources to continue the simulations at high resolution until a significant injection of gas occurred. Note the lack of growth in Accretion Event 4 with a $1000\,M_\odot$ BH after 0.5 Myr of evolution; the absence of a disc makes it vulnerable to the destructive effects of feedback. We prioritized the comparison of thermal feedback parameters under more favourable conditions to obtain a reasonable distribution of behaviour with reasonable feedback efficiencies. Nevertheless, there may be particular formation channels that could lead to such a BH forming, as discussed in Section 1.

We have not accounted for an evolving BH spin. Instead, we have implicitly assumed a maximum fixed spin of about $a = 0.3$ for $\eta = 0.1$. This is the standard solution for a Schwarzschild BH in the thin-disc model, though the radiative efficiency can quadruple for highly spinning BHs (Blandford & Znajek 1977; Tchekhovskoy, Narayan & McKinney 2011). Madau, Haardt & Dotti (2014) argue that the assumptions of the thin-disc model break down at accretion rates where $f_{Edd} \geq 0.3$ and that a slim-disc solution is more appropriate in this regime. As a result of photon trapping, the efficiency of converting gravitational energy into radiative flux in slim discs decreases with increasing accretion rate. Because of low radiative efficiencies around static as well as rapidly rotating holes, the mass growth in the mildly super-Eddington regime becomes nearly independent of the spin parameter. This is in contrast to the thin-disc solution, where the mass of the growing hole is exponentially sensitive to its spin (see fig. 2 of Madau et al. 2014). Therefore, if spin were assigned to the BHs discussed in Section 3.4 that spend all their time at trans- or super-Eddington modes, it would make little difference to the mass growth. However, simulations that spend considerable time in regimes $f_{Edd} < 0.3$ would be limited even further, widening the gap in accretion behaviour of the BHs studied in this work. The spin of the BH is usually assigned based on its formation mechanism; direct-collapse BHs are thought to have lower spin due to strong radial inflows and lack of angular momentum required to form them, while strongly spinning BHs may be more prevalent in the low-mass population due to inheriting the spin of its progenitor star and fewer mergers (Bustamante & Springel 2019; Beckmann et al. 2024). We defer a more self-consistent treatment of spin to later work.

## 6 CONCLUSIONS

We investigated the impact of thermal feedback on super-Eddington accretion in early Universe BHs using sub-pc resolution cosmological simulations. We allowed a single stellar-mass BH born in a dark matter mini-halo to evolve unencumbered by feedback for a certain period of time, and then resimulated with thermal feedback activated before a significant accretion event. We did this for four accretion events, resulting in four simulations with initial BH masses ranging from $10^3$ to $10^5\,M_\odot$. Feedback was modelled thermally, injecting energy into the gas surrounding BHs according to accretion rates and radiative efficiencies ($\epsilon_r$), while coupling efficiencies ($\epsilon_f$) and feedback sphere radii ($r_{fb}$) were varied systematically. We incorporated a detailed chemical network for primordial gas and analysed gas clump formation, fragmentation, and the resulting effects on accretion. Additionally, we examined how feedback efficiency modulates gas expulsion, disc stability, and clump survival. Our key results can be summarized as follows:

(i) The resolution-dependent onset of fragmentation within the nuclear disc around the BH in the absence of feedback enhances the accretion rate.

(ii) Discs with a clumpy, inhomogeneous composition are more resistant to thermal feedback than smooth, homogeneous discs, and feedback destroys smaller clumps first.

(iii) Super-Eddington accretion can be sustained in clumpy environments with feedback efficiencies up to $\epsilon_r = 0.01$ and $\epsilon_f = 0.01$ when the feedback injection radius is $r_{fb} = 5\,dx$, or at higher efficiencies $\epsilon_r = 0.1$ and $\epsilon_f = 0.05$ if the injection radius is increased to $r_{fb} = 10\,dx$.

(iv) Super-Eddington accretion is responsible for the majority of the mass growth in most simulations, even when the BH spends less than 10 per cent of the time in this mode.

(v) Correlations exist between the relative density enclosed in the disc, the relative density enclosed in the feedback region, and the thermal feedback parameters. These relationships can be leveraged to help determine whether feedback has impacted the environment significantly.

(vi) We classify three simulations that accrete at trans-Eddington rates to do so in the presence of effective feedback.

(vii) Two other simulations may also have effective feedback, both of which accrete at twice the Eddington rate.

Determining whether super-Eddington accretion is feasible at cosmic dawn is crucial for investigating whether such a regime could explain the rapid growth of BHs to the billion-solar-mass scale observed in high-redshift quasars. Our results indicate that, in a gas-rich environment, trans-Eddington and super-Eddington accretion are achievable during the early phases of evolution of the first generation of BHs, even in the presence of effective thermal feedback.

## ACKNOWLEDGEMENTS

We would like to thank Alessandro Lupi for providing the observational data with references in Fig. 9. We also extend our gratitude to Molly Peeples for her generous provision of computing resources that enabled this work to be completed. Finally, we thank the anonymous referee for their insightful feedback, which considerably enhanced the quality of this work. STG was supported by the Science and Technologies Facilities Council (STFC) PhD studentship. BDS and SK were supported by the STFC Consolidated Grant RA5496. SK acknowledges funding via STFC Small Grant ST/Y001133/1. RSB acknowledges support by a UKRI future leaders fellowship under grant number MR/Y015517/1. The simulations were run on the high-performance computing (HPC) facility Pleiades. Pleiades is a distributed-memory SGI/HPE ICE cluster, part of the NASA High-End Computing (HEC) Program through the NASA Advanced Supercomputing (NAS) Division at Ames Research Center. Computations and associated analysis described in this work were performed using the publicly available ENZO code (Bryan & Enzo Collaboration 2014) and the YT (Turk et al. 2011) analysis toolkit, which are the products of collaborative efforts of many independent scientists from institutions around the world. For the purpose of open access, the authors have applied a Creative Commons Attribution (CC BY) licence to any Author Accepted Manuscript version arising from this submission.

## DATA AVAILABILITY

All simulations in this work were run from a GitHub fork of ENZO (Bryan & Enzo Collaboration 2014) containing changes, which, at the time of writing, have not been merged into the main branch. Details of the branch used are available on request from the author. All plots were produced using the YTPYTHON package (Turk et al. 2011) and the scripts are available in the GitHub repository.[2]

---

[2] https://github.com/simonetgordon/smartstar_simulation_analysis

# APPENDIX A: EFFECTIVE FEEDBACK SUPPLEMENTAL ANALYSIS

To supplement the classification of simulations as having effective or ineffective thermal feedback in Section 3.5, we provide cumulative mass enclosed radial profiles and projections of temperature and density for Event 1 and Event 3 simulations. Fig. A1 pertains to Event 1, and the projections illustrate 1L16-$\epsilon_r$0.01-$\epsilon_f$0.001-5dx in Fig. A1(b) and 1L16-no-feedback in Fig. A1(c). This low feedback efficiency simulation exhibits few differences in temperature distribution with the no-feedback run, though it does have a disrupted outer disc structure. It also encloses less mass out to $r = 1.20$ pc (blue line) in comparison to its initial state (green line). It occupies a mid-way point between the no-feedback run and the maximum feedback run 1L16-$\epsilon_r$0.1-$\epsilon_f$0.05-5dx (crimson line). It is more deviant from the no-feedback run than its earlier starting counterpart, 1E16-$\epsilon_r$0.01-$\epsilon_f$0.001-5dx (cyan line), and also accretes double the mass in the same period of time. This is due in part to its evolution coinciding with the accretion event, rather than stopping just before. This illustrates that weak feedback can still affect the environment when it is triggered during a strong inflow of material. However, we still categorize 1L16-$\epsilon_r$0.01-$\epsilon_f$0.001-5dx as not generating effective feedback as the relative density in the feedback region is mid-high at about 0.05 and the temperature at pc scales is still $T < 1000$ K.

Fig. A2 pertains to Event 3, and the projections illustrate 3L16-$\epsilon_r$0.01-$\epsilon_f$0.01-5dx in Fig. A2(b) at an edge-on perspective and Fig. A2(c) at a face-on perspective. Unlike 1L16-$\epsilon_r$0.01-$\epsilon_f$0.001-5dx, this simulation does heat its environment to warm temperatures of $T > 10^4$ K. However, the sub-pc-scale nuclear disc remains intact, though with a hot and underdense 'hole' around the BH about the size of the feedback region. Due to its low relative density in the feedback region and very low relative density enclosed within the environment of $<0.1$ pc (the feedback depletes the gas in the inner disc) in Fig. 8, we conclude that feedback is indeed effective in this simulation.

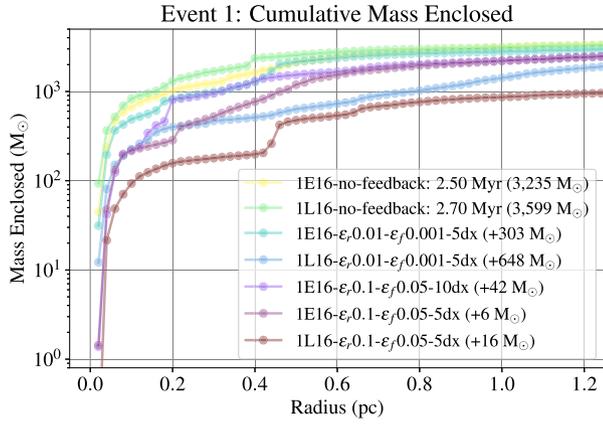

(a) The cumulative mass enclosed around black holes in Event 1 simulations under various thermal feedback scenarios. There are two initial states: one at 2.50 Myr (yellow) and one at 2.70 Myr (green). The other simulations are measured 200 kyr after these points in the presence of feedback.

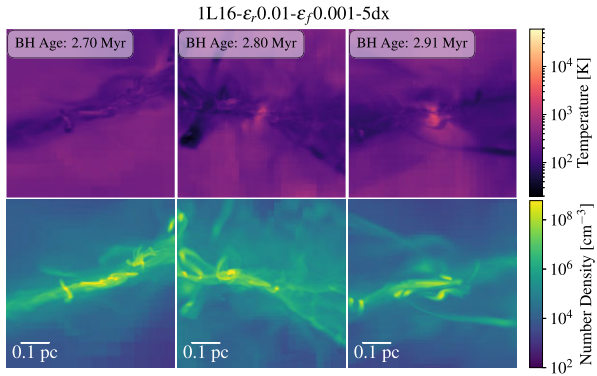

(b) Projections of temperature and density of the region around the black hole in 1L16-$\epsilon_r$0.01-$\epsilon_f$0.001-5dx from an edge-on perspective of the disc.

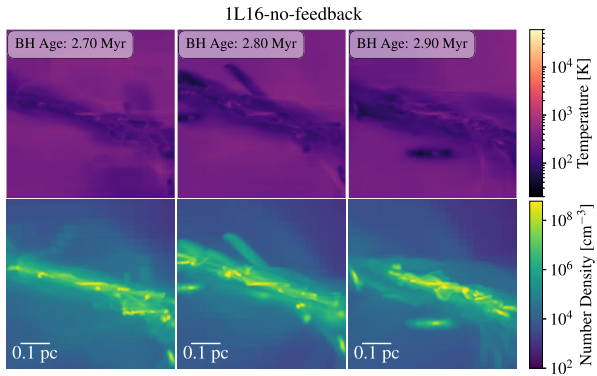

(c) Projections of temperature and density of the region around the black hole in 1L16-no-feedback from an edge-on perspective of the disc.

**Figure A1.** Comparison of cumulative mass enclosed around BHs in Event 1 (a) and projections of density and temperature edge-on to the disc in 1L16-$\epsilon_r$0.01-$\epsilon_f$0.001-5dx (b) and 1L16-no-feedback (c). This low feedback efficiency simulation has a very similar mass profile to the fiducial no-feedback 'late' run at $t = 2.70$ Myr. This illustrates ineffective thermal feedback.

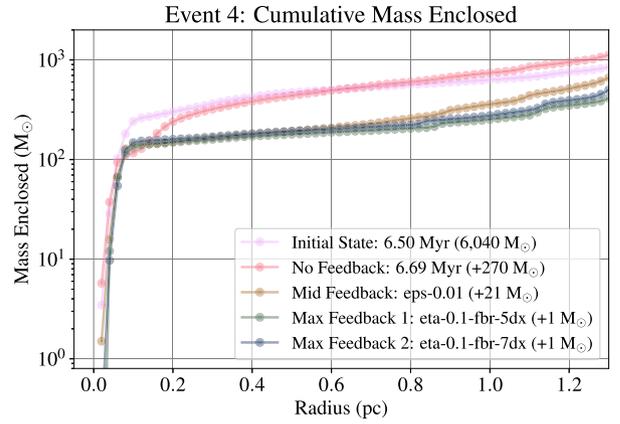

(a) The cumulative mass enclosed around black holes in Event 4 simulations under various thermal feedback scenarios. The initial state at 6.50 Myr (yellow) serves as a reference, with the black hole mass at $6,040\,M_\odot$. The other three simulations are measured at 6.70 Myr, after 200 kyr of evolution in the presence of feedback. The value in brackets indicates the mass gained over this period.

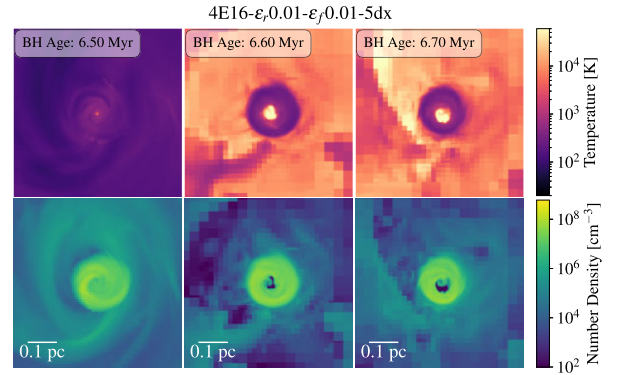

(b) Projections of temperature and density of the region around the black hole from an edge-on perspective of the disc.

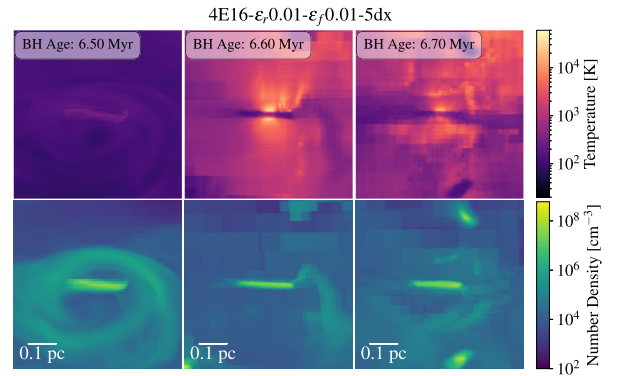

(c) Projections of temperature and density of the region around the black hole from a face-on perspective of the disc.

**Figure A2.** Comparison of cumulative mass enclosed around BHs in Event 4 (a) and projections of density and temperature edge-on (b) and face-on (c) to the disc.

This paper has been typeset from a T<sub>E</sub>X/LAT<sub>E</sub>X file prepared by the author.